\documentstyle[prl,aps,floats,epsfig]{revtex}

\newcommand{\micron}{$\mu$m}

\newcommand{\jpsi}{$J/\psi$}

\newcommand{\ks}{$K_{S}$}

\newcommand{\ksta}{$K^{*0}$}

\newcommand{\bjpsiks}{$B^{0}\to J/\psi K_{S}$}

\newcommand{\bjpsikstar}{$B^{0}\to J/\psi K^{*0}$}

\newcommand{\bpsipks}{$B^{0}\to \psi(2S) K_{S}$}

\newcommand{\bchiciks}{$B^{0}\to \chi_{c1} K_{S}$}

\newcommand{\dz}{$\Delta z$}

\newcommand{\dslnu}{$D^{*} \ell \nu$}

\newcommand{\psipjpsipp}{$\psi (2S) \to J/\psi \pi^+ \pi^-$}

\def\bzb{{\overline{B}{}^0}}

\newcommand{\etal}{\it et al.\rm}

\draft 

\begin{document}

\epsfysize3cm
\epsfbox{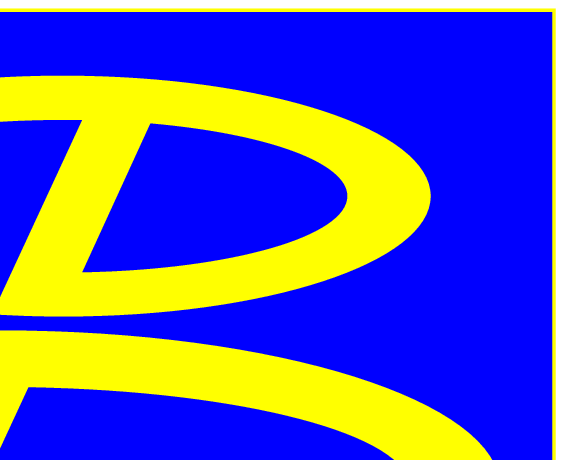}
\vskip -3cm
\hspace*{5.5in}{KEK Preprint 2001-172}\\
\vskip -3mm
\hspace*{5.5in}{Belle Preprint 2002-6}

\begin{center}
\vskip 3cm
{\Large
Observation of Mixing-induced $CP$ Violation in the Neutral $B$ Meson System}
\end{center}

\vskip 1cm
\begin{center}
{\Large{The Belle Collaboration}}
\vskip 1em
  K.~Abe$^{9}$,               
  K.~Abe$^{43}$,              
  R.~Abe$^{31}$,              
  T.~Abe$^{44}$,              
  I.~Adachi$^{9}$,            
  Byoung~Sup~Ahn$^{17}$,      
  H.~Aihara$^{45}$,           
  M.~Akatsu$^{24}$,           
  Y.~Asano$^{50}$,            
  T.~Aso$^{49}$
% ...
\\

\end{center}

\small
\begin{center}
% ...
$^{9}${High Energy Accelerator Research Organization (KEK), Tsukuba}\\
% ...
$^{17}${Korea University, Seoul}\\
% ...
$^{24}${Nagoya University, Nagoya}\\
% ...
$^{31}${Niigata University, Niigata}\\
% ...
$^{43}${Tohoku Gakuin University, Tagajo}\\
$^{44}${Tohoku University, Sendai}\\
$^{45}${University of Tokyo, Tokyo}\\
% ...
$^{49}${Toyama National College of Maritime Technology, Toyama}\\
$^{50}${University of Tsukuba, Tsukuba}\\
% ...
\end{center}
\normalsize

\tighten

\begin{abstract}
This report describes an observation of mixing-induced $CP$ violation 
and a measurement of the $CP$ violation parameter, $\sin 2\phi_1$, 
with the Belle detector at the KEKB asymmetric $e^+e^-$ collider.
Using a data sample of 29.1~fb$^{-1}$ 
recorded on the $\Upsilon(4S)$ resonance that contains 
31.3 million $B\overline{B}$ pairs,
we reconstruct decays of neutral $B$ mesons to the following
$CP$ eigenstates:
$J/\psi K_S^0$, $\psi(2S)K_S^0$, $\chi_{c1}K_S^0$, $\eta_c K_S^0$,
$J/\psi K_L^0$ and $J/\psi K^{*0}$.
The flavor of the accompanying $B$ meson is identified
by combining information from
primary and secondary leptons, $K^\pm$ mesons, $\Lambda$ baryons, slow and fast pions.
The proper-time interval between the two $B$ meson decays
is determined from the distance
between the two decay vertices measured with a silicon vertex detector.

The result
$\sin 2\phi_1 = 0.99 \pm 0.14({\rm stat}) \pm 0.06({\rm syst})$
is obtained by applying
a maximum likelihood fit to the 1137 candidate events.
We conclude that there is large $CP$
violation in the neutral $B$ meson system.  
A zero value for
$\sin 2\phi_1$ is ruled out by more than six standard deviations.
\vskip1pc
\end{abstract}

{\renewcommand{\thefootnote}{\fnsymbol{footnote}}

\setcounter{footnote}{0}
\newpage

\def\PCP{{\cal P}_{CP}}

\twocolumn

\section{Introduction}
The phenomenon of $CP$ violation is one of the major unresolved issues 
in our understanding of particle physics today. 
In 1973, Kobayashi and Maskawa (KM) proposed a model where
$CP$ violation is accommodated as an irreducible complex
phase in the weak-interaction quark mixing matrix\cite{KMpaper},
which is defined as
\begin{equation}
\left( \begin{array}{ccc}
V_{ud} & V_{us} & V_{ub} \\
V_{cd} & V_{cs} & V_{cb} \\
V_{td} & V_{ts} & V_{tb} 
\end{array}\right),
\end{equation}
where
the nontrivial complex phases are conventionally assigned to the furthest
off-diagonal elements $V_{ub}$ and $V_{td}$.  
Unitarity of this CKM matrix (Cabibbo-Kobayashi-Maskawa matrix)
implies that $\Sigma_i V_{ij} V_{ik}^*=\delta_{jk}$, which gives the following relation involving $V_{ub}$ and $V_{td}$:
\begin{equation}
V_{ud}V_{ub}^*+V_{cd}V_{cb}^*+V_{td}V_{tb}^*=0.
\end{equation} 
This expression can be visualized as
a closed triangle in the 
complex plane as shown in Fig.~\ref{fig:triangle}.
\begin{figure}[!htbp]
\begin{center}
\resizebox{0.48\textwidth}{!}{\includegraphics{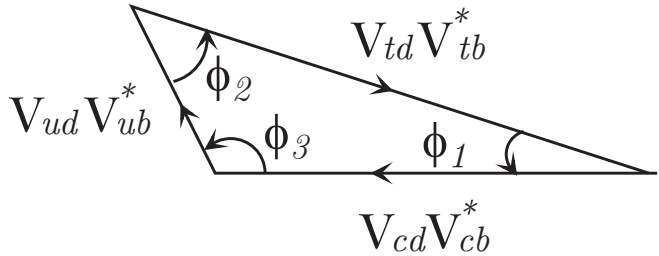}}
\vspace{1ex}
\caption{The unitarity triangle relevant to $B$ decays. Angles of
the triangle arise from the KM phase.} 
\label{fig:triangle}
\end{center}
\end{figure} 
The three interior angles of the unitarity triangle
originate from
the non-vanishing $CP$-violating phase (the KM phase)
and are defined as\cite{BCP}:
\begin{eqnarray}
\phi_1 & \equiv & \pi - 
            {\rm arg}({-V_{td}V_{tb}^*}/{-V_{cd}V_{cb}^*}), \nonumber \\
\phi_2 & \equiv & {\rm arg}({V_{td}V_{tb}^*}/{-V_{ud}V_{ub}^*}), \\
\phi_3 & \equiv & {\rm arg}({V_{ud}V_{ub}^*}/{-V_{cd}V_{cb}^*}). \nonumber
\end{eqnarray}
The bold ansatz of Kobayashi and Maskawa required the existence of six quarks
at a time when
only the $u$, $d$ and $s$ quarks were
known. The subsequent discoveries of the $c$, $b$ and $t$ quarks as well as
the consistency with the $CP$ violation observed in the neutral kaon system
led to the incorporation of the KM mechanism as an essential component
of the Standard Model (SM).

In 1980, Sanda, Bigi and Carter pointed out that the KM model 
contained the possibility of 
sizable $CP$-violating asymmetries in certain neutral $B$
meson decays~\cite{Sanda}. 
The subsequent observation of a long $b$ quark lifetime\cite{MAC}
and a large
mixing in the neutral $B$ meson system\cite{ARGUS} 
indicated that it would be feasible
to measure $CP$ violation in $B$ meson decays at an asymmetric $e^+e^-$ collider
at the $\Upsilon(4S)$ energy.

Until recently the only observation of $CP$ violation
was in the neutral kaon system, where the interpretation of results is complicated
due to large corrections from the strong interaction. By contrast,
these corrections are absent or very small for the aforementioned
$CP$ violation in the neutral $B$ meson system.
Thus its measurement can be used to over-constrain and test the consistency
of the SM.

A pair of neutral $B$ mesons created in
the decay $\Upsilon(4S) \rightarrow B^0 \bzb$ is in a 
state with $C = -1$
at the time of production ($t=0$), where $C$ denotes the charge conjugation.
Although oscillation then starts,
the state preserves the $C$ odd configuration
and is not allowed to be $B^0B^0$ or $\bzb\bzb$.
The time evolution of the pair is given by
\begin{eqnarray}
|\Psi(t)\rangle & = & e^{-t/\tau_{B^0}}|\Psi(t=0)\rangle,\\
|\Psi(t=0)\rangle & = &
 \frac{1}{\sqrt{2}}
 \left [ |B^0(\vec{k})\bzb(-\vec{k})\rangle-
 |B^0(-\vec{k})\bzb(\vec{k})\rangle
 \right], \nonumber
\end{eqnarray}
where $\vec{k}$ and $-\vec{k}$ are the $B$ mesons' momenta
in the $\Upsilon(4S)$ rest frame.
This coherence is preserved until one $B$ meson decays.
Hence, if we can determine the flavor and the decay time $t_{\rm tag}$ of one
of the $B$ mesons decaying into a final state $f_{\rm tag}$,
we are able to determine the time-dependent decay amplitude of the other $B$
at any time $t$ as a function of the time difference $t - t_{\rm tag}$. 
We consider the case where the other $B$ meson
decays at $t = t_{CP}$ to a $CP$ eigenstate, $f_{CP}$.
When the decay is dominated by a single transition amplitude,
the following formulae for the decay rates hold to a good approximation\cite{Sanda}:
\begin{eqnarray}
R(B^0\rightarrow f_{CP};\;\Delta t) &=&
  e^{-|\Delta t|/\tau_{B^0}}/2\tau_{B^0}
\nonumber \\
 & \times& [ 1 + \xi_f\sin 2\phi_{CP}\sin(\Delta m_d \Delta t) ],
\nonumber \\
R(\bzb\rightarrow f_{CP};\;\Delta t) &=&
  e^{-|\Delta t|/\tau_{B^0}}/2\tau_{B^0}
\\
& \times& [ 1 - \xi_f\sin 2\phi_{CP}\sin(\Delta m_d \Delta t) ],
\nonumber
\end{eqnarray}
and the time-dependent
$CP$-violating asymmetry is\cite{PDG2000}
\begin{eqnarray}
A(\Delta t) &\equiv& 
  \frac{R(\bzb\rightarrow f_{CP};\;\Delta t)-R(B^0\rightarrow f_{CP};\;\Delta t)}
       {R(\bzb\rightarrow f_{CP};\;\Delta t)+R(B^0\rightarrow f_{CP};\;\Delta t)}
 \nonumber \\
 &=& -\xi_f\sin 2\phi_{CP}\sin(\Delta m_d \Delta t),
\end{eqnarray}
where $\xi_f$ is the $CP$ eigenvalue of $f_{CP}$, 
$\Delta m_d$ is the mass difference 
between the two $B^0$ meson mass eigenstates\cite{signDeltaM}
and $\Delta t = t_{CP} -t_{\rm tag}$.
Because the asymmetry, $A(\Delta t)$, vanishes in the time-integrated rate,
it is very important to measure the time dependence.

The angle $\phi_{CP}$ is
directly related to the interior angles of the unitarity triangle, and
is the phase difference between two interfering amplitudes, one
for $B^0(\bzb) \rightarrow f_{CP}$ and the other
for the mixing process $B^0(\bzb) \rightarrow \bzb(B^0) 
\rightarrow f_{CP}$.
The quantity
$\phi_{CP}$ is equal to $\phi_1$ if $f_{CP} = J/\psi K_S^0$ or any other $CP$ eigenstate that
arises from a $b(\overline{b})\rightarrow c{\overline c}s(\overline{s})$ transition. 
The hadronic uncertainty in this case is negligibly small
because the amplitude
of the $b \rightarrow s$ flavor-changing transition
with associated $c\overline{c}$ production
is not only small but has the same weak phase. 

As is described in the next section,
the KEKB $e^+ e^-$ collider produces 
the $\Upsilon(4S)$ with a Lorentz boost of
$\beta \gamma = 0.425$. 
Since the $B^0$ and $\bzb$ are nearly at rest in the
$\Upsilon(4S)$ center of mass system (cms), $\Delta t$ can be
determined from the displacement 
between the two $B$ decay vertices---{\it i.e.}
\begin{equation}
\Delta t \simeq (z_{CP} - z_{\rm tag})/\beta\gamma c
 \equiv \Delta z/\beta\gamma c,
\end{equation}
where the $z$ axis is defined to be anti-parallel to the positron beam direction.

Following initial experimental studies\cite{Early_cp}\cite{BELLE_cp1},
the BaBar\cite{BABAR_cp2} and Belle\cite{BELLE_cp2} Collaborations recently
reported the first clear observations of $CP$ violation in the neutral $B$ meson system.
In this paper, 
we describe the details of the measurement of $\sin 2\phi_1$ with
the Belle detector at the KEKB asymmetric $e^+e^-$ collider with the
same 29.1~fb$^{-1}$ data sample reported in Ref.\cite{BELLE_cp2}. 
In the next section we describe the KEKB collider and the Belle detector.
The measurement of $\sin 2\phi_1$  
requires the reconstruction of $B^0\to f_{CP}$
decays (denoted by $B_{CP}$), 
the determination of the $b$-flavor of the accompanying (tagging)
$B$ meson, the measurement of $\Delta t$,
and a fit of the expected $\Delta t$ distribution to 
the measured distribution using a likelihood method.
The selection and tagging procedures are described in 
Sections~\ref{sec:rec_bcp} and \ref{sec:flavor_tagging}.
After introducing the methods to extract $\sin 2\phi_1$ 
from the $\Delta t$ distributions in Section~\ref{sec:maximum_likelihood_fit},
we present the results of the fit 
and discuss the interpretation of the $CP$ violation
in Section~\ref{sec:fit_results}. We summarize the results in
Section~\ref{sec:conclusion}.

\section{Experimental apparatus}
\label{sec:experimental_apparatus}

KEKB\cite{KEKB} is an asymmetric $e^+e^-$ collider 3~km in circumference,
which consists of 8~GeV $e^-$ and 3.5~GeV $e^+$ storage rings and an injection
linear accelerator. 
It has a single interaction point (IP) where the $e^+$ and
$e^-$ collide with a crossing angle of 22~mrad. 
The data used in this analysis were taken between January 2000 and July 2001.
The collider was operated during this period
with a peak beam current of 930~mA($e^+$) and 
780~mA($e^-$), giving a peak luminosity of 4.5$\times 10^{33}$~cm$^{-2}$s$^{-1}$. 
Due to the energy asymmetry, the $\Upsilon(4S)$ resonance and its daughter $B$ mesons are
produced with a Lorentz boost of
$\beta\gamma =$0.425.
On average,
the $B$ mesons decay approximately
200~$\mu$m from the $\Upsilon(4S)$ production point.

The Belle detector~\cite{Belle} is a general-purpose large solid 
angle magnetic spectrometer surrounding the interaction point.
It consists of a barrel, forward and rear components.
It is placed in such a way that the axis of the 
detector solenoid is parallel to the $z$ axis. In this way we
minimize the Lorentz force on the low energy positron beam.

Precision tracking and vertex measurements are provided
by a central drift chamber (CDC)~\cite{CDC} and 
a silicon vertex detector (SVD)~\cite{SVD}.
The CDC is
a small-cell cylindrical drift 
chamber with 50 layers of anode wires including 18 layers of stereo 
wires.  A low-$Z$ gas mixture
(He (50\%) and ${\rm C}_2{\rm H}_6$ (50\%)) is used to minimize multiple Coulomb scattering to
ensure a good momentum resolution, especially for low momentum particles.
It provides three-dimensional trajectories of charged particles
in the polar angle region $17^\circ < \theta < 150^\circ$
in the laboratory frame, where $\theta$ is measured with respect to the $z$ axis.
The SVD consists of three layers of double-sided silicon strip detectors
arranged in a barrel and covers 86\% of the solid angle. The three layers
at radii of 3.0, 4.5 and 6.0 cm
surround the beam-pipe, a double-wall beryllium cylinder 
of 2.3~cm radius and 1~mm  thickness.
The strip pitches are 84~$\mu$m for the measurement of $z$
coordinate and 25~$\mu$m for the measurement of azimuthal angle $\phi$.
The impact parameter resolution for reconstructed tracks is measured as
a function of the track momentum $p$ (measured in GeV/{\it c}) to be 
$\sigma_{xy}$ = [19 $\oplus$ 50/($p\beta\sin^{3/2}\theta$)]~$\mu$m and
$\sigma_{z}$ = [36 $\oplus$ 42/($p\beta\sin^{5/2}\theta $)]~$\mu$m.
The momentum resolution of the combined tracking system is 
$\sigma_{p_{\rm t}}/p_{\rm t} = (0.30/\beta \oplus 0.19p_{\rm t})$\%, 
where $p_{\rm t}$ is the transverse momentum in GeV/{\it c}.

The identification of charged pions and kaons uses three detector systems:
the CDC measurements of $dE/dx$, a set of time-of-flight 
counters (TOF)\cite{TOF} and a set of aerogel Cherenkov counters (ACC)\cite{ACC}. 
The CDC measures 
energy loss for charged particles with a resolution of $\sigma(dE/dx)$ = 6.9\%
for minimum-ionizing pions. 
The TOF consists of 128 plastic scintillators viewed on both ends
by fine-mesh photo-multipliers that operate stably
in the 1.5~T magnetic field. Their time resolution is 95~ps
($rms$) for minimum-ionizing particles, 
providing three standard deviation (3$\sigma$) $K^\pm/\pi^\pm$ separation below 
1.0~GeV/$c$, and 2$\sigma$
up to 1.5~GeV/$c$.
The ACC consists of 1188 aerogel blocks with refractive indices between
1.01 and 1.03 depending on the polar angle. Fine-mesh photo-multipliers detect the Cherenkov light.
The effective number of photoelectrons is approximately 6
for $\beta =1$ particles.
Using this information, $P(K/\pi) = Prob(K)/(Prob(K)+Prob(\pi))$, the
probability for a particle to be a $K^\pm$ meson, is calculated. 
A selection with $P(K/\pi) > 0.6$ retains about 90\% of the charged kaons
with a charged pion misidentification rate of about 6\%. 

Photons and other neutrals are reconstructed in a CsI(Tl) crystal calorimeter
(ECL)~\cite{ECL} consisting of 8736 crystal blocks, 16.1 radiation lengths ($X_0$) thick.
Their energy resolution is 1.8\% for photons above 3 GeV.
The ECL covers the same angular region as the CDC.
Electron identification in Belle is based on
a combination of $dE/dx$ measurements
in the CDC, the response of the ACC, the position and the shape
of the electromagnetic shower, as well as the ratio of the 
cluster energy to the particle momentum\cite{EID}.
The electron identification efficiency is determined 
from two-photon $e^+e^-\rightarrow e^+e^-e^+e^-$ 
processes to be more than 90\% for $p >1.0$~GeV/{\it c}.
The hadron misidentification probability,
determined using tagged pions from
inclusive $K_S^0\rightarrow \pi^+\pi^-$ decays, is below $0.5\%$.  

All the detectors mentioned above are inside
a super-conducting solenoid of 1.7~m radius that generates a 1.5~T magnetic field. The outermost 
spectrometer subsystem is
a $K_L^0$ and muon detector (KLM)\cite{KLM}, which consists of 14 layers
of iron absorber (4.7~cm thick) alternating with resistive plate counters 
(RPC). The KLM system covers polar angles
between 20 and 155 degrees.
The overall muon
identification efficiency, determined by using
a two-photon process 
$e^+e^-\rightarrow e^+e^-\mu^+\mu^-$  
and simulated muons embedded in $B\overline{B}$ candidate events,
is greater than 90\% for tracks with
$p > 1$~GeV/{\it c} detected
in the CDC.  The corresponding
pion misidentification probability, determined using
$K_S^0\rightarrow \pi^+\pi^-$  decays, is less than 2\%. 

In our analysis,
Monte Carlo (MC) events are generated using the $QQ$ event
generator \cite{QQ} and the 
response of the Belle detector is precisely simulated by a
GEANT3-based program\cite{GEANT}.
The simulated events are then reconstructed and analyzed with the
same procedure as is used for the real data.

\section{Reconstruction of {\boldmath $B^0$} decays}
\label{sec:rec_bcp}

We use a 29.1~fb$^{-1}$ data sample, which contains 31.3 million $B \overline B$
pairs, accumulated at the $\Upsilon(4S)$  resonance
between January 2000 and July 2001.
The entire data sample has been analyzed and reconstructed with the
same procedure.

We reconstruct $B^0$ decays to the following ${CP}$ eigenstates~\cite{CC}:
$J/\psi K_S^0$, $\psi(2S)K_S^0$, $\chi_{c1}K_S^0$ and $\eta_c K_S^0$ having $\xi_f=-1$; 
and $J/\psi K_L^0$ having $\xi_f=+1$.
We also use the decay $B^0\to J/\psi K^{*0}$,
$K^{*0}\to  K_S^0\pi^0$, which
is a mixture of even
and odd $CP$ eigenstates.
The selection of these $B_{CP}$ candidates is described in
the following sections.

\subsection{\boldmath $B{\overline B}$ event pre-selection}

To select generic $B{\overline B}$ candidates, 
we require at
least three tracks that satisfy
$\sqrt{x^2+y^2} < 2.0$~cm, $|z| < 4.0$~cm, and $p_{\rm t} > 0.1~{\rm GeV}/c$,
where $x$, $y$, $z$ represent the point
of closest approach of the track to the beam axis,
and $p_{\rm t}$ is 
the momentum of the track projected onto the $xy$-plane.
We also require that 
more than one neutral cluster is observed
and have energy greater than 0.1~GeV.

The sum of all cluster energies,
boosted back to the cms 
assuming each cluster is generated by a  massless
particle, is required to be between 10\% and 80\% of the total cms
energy.
The total visible energy in the cms, $E_{\rm vis}^{\rm cms}$, is computed
from the selected tracks, assuming they are pions,
and the calorimeter clusters that are not associated with the tracks.
We require that $E_{\rm vis}^{\rm cms}$ is greater than 20\% of the total cms energy.
The absolute value of the $z$ component of the
cms momentum is required to
be less than 50\% of the cms energy.
The event vertex reconstructed from the selected tracks 
must be within 1.5~cm and 3.5~cm of the interaction region
in the directions perpendicular and parallel to the $z$ axis, respectively.
Monte Carlo simulation shows that the
selection criteria described above retain more than 99\%
of $B\overline{B}$ events and $J/\psi$ inclusive events.

To suppress continuum background, which consists of $e^+e^- \rightarrow q\bar{q}$
where $q$ is $u$, $d$, $s$ or $c$ quark,
we also require
$R_2 \equiv H_2/H_0 \leq 0.5$, where $H_2$ and $H_0$ are the second and zeroth
Fox-Wolfram moments\cite{FW}.

\subsection{\boldmath $B^0\rightarrow {\rm Charmonium}~K_S^0(K^{*0})$ reconstruction}
The candidate 
$J/\psi$ and $\psi(2S)$ mesons are reconstructed using 
their decays to lepton pairs, i.e.
$J/\psi \rightarrow \mu^+\mu^-$ and $e^+e^-$.
The $\psi(2S)$ meson is also reconstructed via its  $J/\psi\pi^+\pi^-$ decay,
the $\chi_{c1}$ meson via its $J/\psi\gamma$ decay, and
the $\eta_c$ meson via its $K^+K^-\pi^0$
and $K_S^0(\pi^+\pi^-)K^{\pm}\pi^{\mp}$ 
decays.

For $J/\psi$ and $\psi(2S)\to\ell^+\ell^-$ decays, we
use oppositely charged track pairs where both tracks are 
positively identified as leptons.  
For the $B^0 \to J/\psi K_S^0(\pi^+\pi^-)$ mode,
which has the smallest background fraction among 
the $CP$ eigenstates that are used, the 
requirement for {\em one} of the tracks is relaxed
to improve the efficiency: 
a track with an ECL energy deposit consistent with  
a minimum ionizing particle is accepted
as a muon and a track that satisfies 
either the $dE/dx$ or the ECL shower energy requirements 
as an electron.
In order to remove either badly measured tracks or tracks that do not come
from the interaction region, we require $|dz| <$ 5~cm for both 
lepton tracks.
In order to account partially for final-state radiation and
bremsstrahlung, the invariant mass calculation of the $e^+e^-$ pairs
is corrected by adding photons found within 50~mrad of
the $e^+$ or $e^-$ direction.
Nevertheless, a radiative tail remains and we use an
asymmetric invariant mass requirement 
$-150 \le M_{e^+e^-}-M_{J/\psi(\psi(2S))} \le 36 {\rm~MeV}/c^2$. 
Since the $\mu^+\mu^-$ radiative tail is smaller, we select
$-60 \le M_{\mu^+\mu^-} - M_{J/\psi(\psi(2S))} \le  
36 {\rm~MeV}/c^2$\cite{footnoteJPSIKSTAR}.
Events with a candidate $J/\psi \rightarrow \ell^+\ell^-$ decay are 
accepted if the $J/\psi$
momentum in the cms is below 2~GeV/$c$.  
Figure \ref{jpsimass} shows the invariant mass distributions for 
$J/\psi\rightarrow \mu^+\mu^-$ and $J/\psi\rightarrow e^+e^-$ 
with the selection criteria applied for the $J/\psi K_S^0$
mode.

\begin{figure}[!htbp]
\begin{center}
\resizebox{0.48\textwidth}{!}{\includegraphics{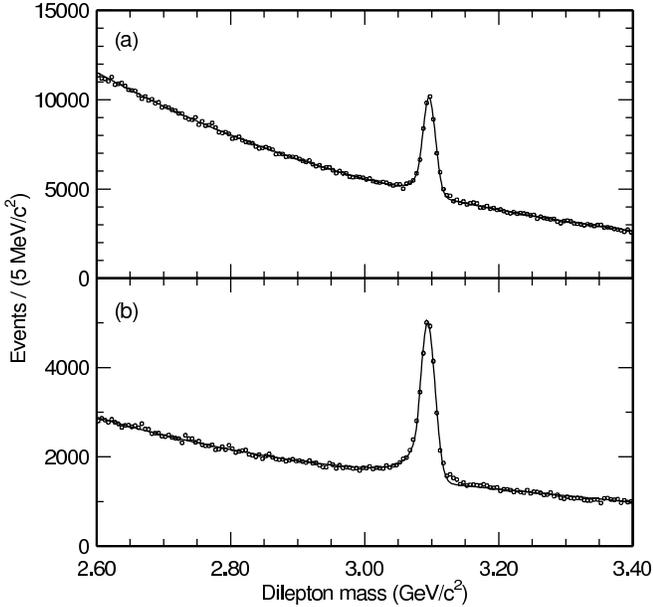}}
\vspace{1ex}
\caption{The invariant mass distributions for (a) $J/\psi\rightarrow
\mu^+\mu^-$ and (b) $J/\psi\rightarrow e^+e^-$ in the reconstruction of the
$J/\psi K_S^0$ mode where the selection
for {\em one} of the tracks is relaxed (details are explained in the
text).}
\label{jpsimass}
\end{center}
\end{figure}

To reconstruct 
$\psi (2S)\rightarrow J/\psi\pi^+\pi^-$ decays, we select $\pi^+\pi^-$ 
pairs with an invariant mass
greater than 400~MeV/{\it c}$^2$.
This requirement is based on the measured
mass distribution\cite{MarkIII} and improves the
signal-to-background ratio.
The $\psi (2S)$
candidates are then selected requiring
the mass difference, $M_{\ell^+\ell^-\pi^+\pi^-}-M_{\ell^+\ell^-}$,
to be between 0.58~GeV/{\it c}$^2$ and 0.60~GeV/{\it c}$^2$.
This corresponds to a $\pm 3\sigma$ requirement where $\sigma$ is the resolution on
the mass difference~\cite{footnoteBRCHARM}.

The $\chi_{c1} \rightarrow J/\psi\gamma$ candidates are selected by
requiring the mass difference, $M_{\ell^+\ell^-\gamma}-M_{\ell^+\ell^-}$,
to be between 0.385~GeV/{\it c}$^2$ and 0.4305~GeV/{\it c}$^2$.
We veto photon candidates that form a good
$\pi^0$ candidate with any other photon
candidate of energy greater than 60 MeV in the event.
A good $\pi^0$ candidate is defined by
an invariant mass
within $-$28 to +17~MeV/{\it c}$^2$ of the nominal $\pi^0$ mass,
and by a $\chi^2$ of less than 10 after a mass-constrained
kinematic fit~\cite{footnoteBRCHARM}.

For $K_S^0 \rightarrow \pi^+\pi^-$ reconstruction, 
we select oppositely charged track pairs 
that satisfy the following requirements:
(1) when both pions have associated SVD hits, 
the distance of closest approach of both pion tracks in the $z$ direction
should be smaller than 1~cm;
(2) when only one of the two pions has associated SVD hits,
the distance of closest approach of both the pion tracks to the nominal interaction point
in the $x$-$y$ plane should be larger than 0.25~mm;
(3) when neither pion has an associated SVD hit,
the $\phi$ coordinate of the  $\pi^+\pi^-$ vertex and the $\phi$
direction of the  $\pi^+\pi^-$ candidate's three momentum vector should agree
within 0.1~radian.
The invariant mass of the candidate $\pi^+\pi^-$ pair is required 
to be between 482 and 514 MeV/$c^2$, which retains 99.7\% of the $K_S^0$ candidates.

For the $\chi_{c1}K_S^0$ and $\eta_c K_S^0$ modes, more stringent track selection criteria are applied in $ K_S^0 \to \pi^+\pi^-$
reconstruction to reduce the background:
(1) the flight length in the $r$-$\phi$ plane should be greater than 
1~mm (2~mm for $\eta_c \to K^+K^-\pi^0$); 
(2) a mismatch in the 
$z$ direction at the $K_S^0$ vertex point for two $\pi^\pm$ tracks should be 
less than 2.5~cm (10~cm for $\eta_c K_S^0$);
(3) the angle in the $r$-$\phi$ plane between the $K_S^0$ momentum
vector and the direction defined by the $K_S^0$ and $J/\psi$ (or $\eta_c$) decay
vertices
should be less than 0.2 (0.1 for $\eta_c K_S^0$) radian; and
(4) for the $\chi_{c1}K_S^0$ selection we also require that
the distance of closest approach of the $J/\psi$ vertex in the radial direction for 
each $\pi^\pm$ track should be greater than 0.25~mm. 

To reconstruct $K_S^0 \rightarrow \pi^0\pi^0$ candidates,
we first select photons that have an energy of at least 20~MeV.
For $\pi^0 \rightarrow \gamma\gamma$ candidates, we require that
the invariant mass of the two photons be
between 80~MeV/{\it c}$^2$ and 150~MeV/{\it c}$^2$ and
the momentum of $\pi^0$ be greater than 100~MeV/{\it c}.
Initially, 
the $\pi^0$ decay vertex is assumed to be at the nominal interaction point.
To select $K_S^0 \rightarrow \pi^0\pi^0$ candidates,
we find the best decay vertex where
the invariant masses of two $\pi^0$ candidates are the most consistent with
the nominal $\pi^0$ mass.
To this end, first the $K_S^0$ flight direction is measured from the 
sum of the momenta of the four 
photons, then we calculate the $\chi^2$ of the mass constrained fit for each 
$\pi^0$, varying the decay vertex along the $K_S^0$ direction through the IP.
We choose the vertex point that minimizes the sum of the $\chi^2$ for the two 
$\pi^0$ candidates.
We require that the distance between the IP and the reconstructed $K_S^0$ decay vertex be 
larger than $-20$~cm
where the positive direction is defined by the $K_S^0$ momentum.
To reduce the combinatorial background, the $\chi^2$ for each $\pi^0$ meson
at the $K_S^0$ decay vertex point determined by this method is also required to be less than 10. 
Using the calculated $K_S^0$ decay vertex,
we finally require that the invariant mass of $K_S^0$ candidate 
lie between 470~MeV/{\it c}$^2$ and 520~MeV/{\it c}$^2$.   

For $J/\psi K^{*0}(K_S^0\pi^0)$ decays, we use $K_S^0\pi^0$ combinations 
that have an invariant mass 
within 75~MeV/$c^2$ of the nominal $K^{*0}$ mass.
Here, the $\pi^0$ candidate is reconstructed from photons with an energy greater than 40 MeV
and the two photon invariant mass is required
to be between 125 and 145 MeV/$c^2$.
We reduce the background from low-momentum $\pi^0$ mesons by
requiring $\cos\theta_{K^*}<0.8$, where $\theta_{K^*}$ is 
the angle between 
the $K^{*0}$ flight direction and
the $K_S^0$  momentum vector
calculated  in the $K^{*0}$ rest frame.

We reconstruct
$B^0 \rightarrow \eta_c K_S^0$ candidates in two $\eta_c$
decay modes: $\eta_c \rightarrow K_S^0 K^\pm \pi^\mp$ 
and $\eta_c \rightarrow K^+ K^- \pi^0$.
We require charged kaons to be positively identified 
using CDC $dE/dx$ measurements and information from the TOF and ACC systems.
For the $\eta_c \rightarrow K_S^0 K^\pm \pi^\mp$ channel,
we require an invariant mass ranging from 2.935 to 3.035~GeV/c$^2$.
In order to suppress the continuum background, we require
$R_2 < 0.45$ and 
$|\cos\theta_{\rm thr}| < 0.85$, where $\theta_{\rm thr}$ is the angle
between the thrust axis of the $B^0$ candidate and that of all remaining
charged and neutral particles in the event.
In the $\eta_c \rightarrow K^+ K^- \pi^0$ mode, we reconstruct the $\pi^0$ meson
from photons having an energy larger than 50 (200) MeV
in the ECL barrel (end-cap) region. 
The invariant mass of the $\eta_c  \rightarrow K^+ K^- \pi^0$ candidate is required to be between
2.890 and 3.040~GeV/$c^2$.
The continuum background is suppressed by requiring 
$R_2 < 0.40$ and $|\cos\theta_{\rm thr}| < 0.80$. 

For $B^0$ reconstruction,  we calculate 
the energy difference, $\Delta E$, and the beam-energy constrained mass, $M_{\rm bc}$.
The energy difference is defined as 
 $\Delta E\equiv E_B^{\rm cms} - E_{\rm beam}^{\rm cms}$
and the beam-energy constrained
mass $M_{\rm bc}\equiv\sqrt{(E_{\rm beam}^{\rm cms})^2-(p_B^{\rm cms})^2}$,
where $E_{\rm beam}^{\rm cms}$ is the cms beam energy,
and $E_B^{\rm cms}$ and $p_B^{\rm cms}$ are the cms energy and momentum
of the $B^0$ candidate.
To improve the momentum resolution,
a vertex fit and then a mass-constrained fit are performed wherever needed.
The resulting fitted momenta are used in the $\Delta E$ and $M_{\rm bc}$ calculations.
A scatter plot of $M_{\rm bc}$ and $\Delta E$ for 
$J/\psi K_S^0(\pi^+\pi^-)$ candidates
is shown
in Fig.~\ref{dEvsMbc_gpm} along with the projections onto each axis.
The $B^0$ candidates are selected by requiring 
5.270 $< M_{\rm bc} <$ 5.290~GeV/$c^2$ ($|M_{\rm bc} - M_{B^0}| < 3.5\sigma$)
and by applying the mode-dependent requirements on $\Delta E$ listed in Table~\ref{Nevents}.
Figure~\ref{mb_all} shows the $M_{\rm bc}$ distribution 
after the $\Delta E$ selection.

The background in the signal region is estimated by simultaneously fitting the
$M_{\rm bc}$ and $\Delta E$ distributions 
with signal and background functions
in the region 5.2 $< M_{\rm bc} <$ 5.3~GeV/$c^2$ and 
$-$0.1 $< \Delta E <$ 0.2~GeV.
We use a two-dimensional Gaussian for the signal.
For the background, we use the ARGUS background function\cite{argus_func} 
for $M_{\rm bc}$ and a linear function for $\Delta E$. 
The details are described in Section ~\ref{sec:signalprob}.
The number of candidates observed as well as the estimated background are given in
Table~\ref{Nevents}.

\begin{figure}[!htb]
\begin{center}
\resizebox{0.48\textwidth}{!}{\includegraphics{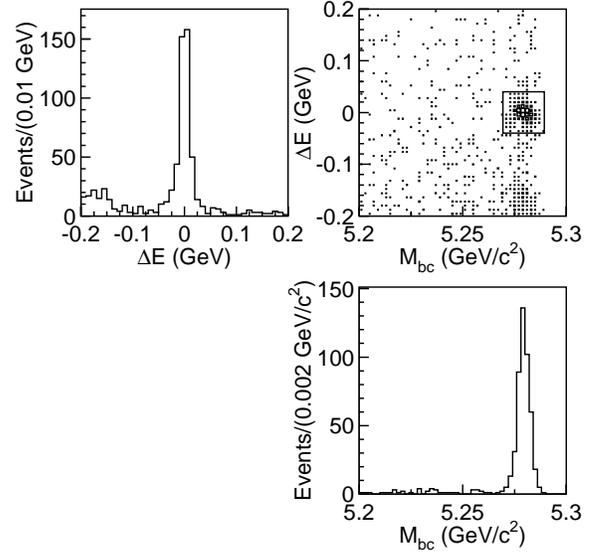}}
\vspace{1ex}
\caption{The scatter plot of $\Delta E$ versus
$M_{\rm bc}$ for $J/\psi K_S^0(\pi^+\pi^-)$ candidates.
The box represents the signal region.
The upper left figure is 
the $\Delta E$ projection with  
5.270 $< M_{\rm bc} <$ 5.290 GeV/$c^2$. 
The lower right figure is the 
$M_{\rm bc}$ projection with $|\Delta E| < $0.04 GeV. The enhancement in the negative
$\Delta E$ region is due to decay modes with additional pions, e.g. 
$B \rightarrow J/\psi K^*$, $K^* \rightarrow K_S^0 \pi$.}
\label{dEvsMbc_gpm}
\end{center}
\end{figure}

\begin{figure}[!htb]
\begin{center}
\resizebox{0.48\textwidth}{!}{\includegraphics{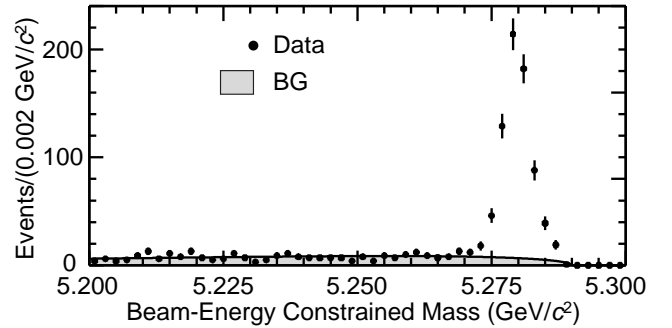}}
\vspace{1ex}
\caption{ The $M_{\rm bc}$ distribution for all exclusive decay modes
combined except $J/\psi K_L^0$.  The $\Delta E$ cuts are imposed.
The shaded area is the estimated background. }
\label{mb_all}
\end{center}
\end{figure}

\begin{table}[htb]
\caption{Summary of $\Delta E$ signal region, the numbers of signal candidates, 
 and expected background. } 
\medskip
\label{Nevents} 
\begin{tabular}{lcccc}
 Decay       & $\Delta E$ cut (MeV) & Signal     & Expected   \\
 mode        & Lower $\sim$ Upper   & candidates & background \\ \hline
$B^0 \rightarrow J/\psi K_S^0$ & & & \\
 $~~K_S^0 \rightarrow \pi^+\pi^- $
& $-$40 $\sim$ 40 & 457  &  11.9    \\  
 $~~K_S^0 \rightarrow \pi^0\pi^0 $
& $-$150 $\sim$ 100 & 76  &  9.4    \\
$B^0 \rightarrow \psi (2S)K_S^0$ & & & \\
 $~~\psi (2S) \rightarrow \ell^+\ell^- $
& $-$40 $\sim$ 40 & 39  &  1.2   \\
 $~~\psi (2S) \rightarrow J/\psi \pi^+\pi^-  $
& $-$40 $\sim$ 40 & 46  &  2.1   \\
$B^0 \rightarrow \chi_{c1}K_S^0$ 
& $-$40 $\sim$ 40 & 24  &  2.4    \\
$B^0 \rightarrow \eta_c K_S^0$ & & & \\
 $~~\eta_c \rightarrow K_S^0 K^\pm \pi^\mp$
& $-$40 $\sim$ 40 & 23  &  11.3    \\
 $~~\eta_c \rightarrow K^+ K^- \pi^0$
& $-$60 $\sim$ 40 & 41  &  13.6    \\
$B^0 \rightarrow J/\psi K^{*0}$ 
& $-$50 $\sim$ 30 & 41  &  6.7    \\
\end{tabular}
\end{table}

\subsection{\boldmath $B^0\rightarrow J/\psi K_L^0$ reconstruction}
\label{sec:psikl_rec}
The reconstruction of $B^0\rightarrow J/\psi K_L^0$ is an experimental challenge
but is very important because its yield is expected to be large.
In addition, since this mode is a $CP$-even eigenstate, we should observe
a time-dependent asymmetry reversed in sign compared to
$J/\psi K_S^0$, which provides an important experimental consistency check.

While the detached vertex and invariant mass of the 
$K_S^0$ provide significant background reduction for
$J/\psi K_S^0$,
the background is larger for $J/\psi K_L^0$
as only the $K_L^0$ direction is measured. 
Since the energy of the $K_L^0$ is not measured, 
$M_{\rm bc}$ and $\Delta E$ cannot be used as the final kinematical variables
to identify $B^0$ candidates as in other final states.
Using the four-momentum of a reconstructed $J/\psi$ candidate 
and the $K_L^0$ flight direction, we calculate the momentum of
the $K_L^0$ candidate requiring $\Delta E = 0$.
We then calculate $p^{\rm cms}_B$ which is used for the final selection.

The selection criteria are necessarily 
tighter than those used for the $J/\psi K_S^0$ candidates.
However, precise determination of the $K_L^0$ flight direction
with the KLM and ECL allows us to reconstruct $J/\psi K_L^0$ candidates
with sufficient efficiency and purity.

We use tracks which are
positively identified as electrons (muons) 
in the identification of $J/\psi \rightarrow e^+e^-~(\mu^+\mu^-)$ candidates.
We require the invariant mass of the lepton pair to lie in the range
3.05 $< M_{\ell^+\ell^-} <$ 3.13 GeV/$c^2$.
The radiative photon correction for electron pairs
is made in the same way as for the other modes.
Events are rejected if one of the following decay modes are exclusively
reconstructed, and
satisfy $|\Delta E| < 0.05$ GeV and 5.27 $< M_{\rm bc} < 5.29$ GeV/$c^2$:
$J/\psi K^+$,
$J/\psi K_S^0$,
$J/\psi K^{*+}$ $(K^{*+} \rightarrow K^+ \pi^0, K_S^0 \pi^+)$,  and
$J/\psi K^{*0}$ $(K^{*0} \rightarrow K^+ \pi^-, K_S^0 \pi^0)$.

We select $K_L^0$ candidates based on the KLM and ECL information.
There are two classes of $K_L^0$ candidates that we refer to as KLM
and ECL candidates. To select the KLM candidates,
a cluster of KLM hits is formed by combining the hits within a $5^{\circ}$ opening angle.
We require hits in two or more KLM layers and calculate
the center of the KLM cluster.
If there is an ECL cluster with energy greater than 0.16 GeV within a 15$^\circ$ cone,
we relax our criteria to allow a cluster with a hit in just one KLM layer.
In this case
the direction of the ECL cluster is taken 
as the $K_L^0$ direction.
If the cluster lies within a 15$^\circ$ cone of the extrapolation
of a charged track to the first layer of the KLM, it is discarded.

ECL candidates are selected from ECL clusters using the following
information:
      the distance between the ECL cluster and the closest charged
      track position;
      the ECL cluster energy;
      the ratio of energies summed in
      3~$\times$~3  and 5~$\times$~5 arrays of CsI crystals 
      surrounding the crystal at the center of the shower;
      the ECL shower width 
      and the invariant mass of the shower. 
After a very loose pre-selection based on the above five
discriminants, we calculate signal and background likelihood
values for each discriminant based on a $J/\psi$ inclusive MC.
Taking the products of the above five likelihoods for each signal
and background, we form the likelihood ratio
${\cal L}_{K_L^0} / ({\cal L}_{K_L^0} + {\cal L}_{\rm fake})$.
This likelihood ratio is required to be greater than 0.5.

We examine the characteristics of $K_L^0$ candidates in both the data
and the MC. We obtain consistent distributions for
the number of $K_L^0$ candidates per event, the $K_L^0$ flight directions
in the laboratory system, the total number of hit KLM layers and
the number of hit first-layers 
in the $K_L^0$ candidates.
We investigate the momentum dependence of the KLM response
by using charged pions and kaons. The MC simulation reproduces
the results obtained with data well.
We also use
$e^+ e^- \to \gamma \phi$ followed by $\phi \to K_L^0 K_S^0$, where
the exact $K_L^0$ direction and momentum can be obtained
by reconstructing $\gamma$ and $K_S^0 \to \pi^+ \pi^-$. These studies
indicate that the $K_L^0$ identification is well reproduced 
by the MC with an exception of an overall detection efficiency. 
The $K_L^0$ detection efficiency in the data is found to be 
lower than the MC expectation.
This, however, does not cause a difficulty in our analysis 
since we do not rely on the MC $K_L^0$ detection efficiency.

For both KLM and ECL candidates, we require that the
$K_L^0$ direction should be within 45$^\circ$ of its expected 
direction calculated from the $J/\psi$ candidate momentum
assuming that the $B^0$ candidate was at rest in the cms.
We also require that
no photon from a $\pi^0$ decay is found near the $K_L^0$ candidate.
For this requirement, we select $\pi^0$ candidates satisfying
$0.12 < M_{\gamma\gamma} < 0.15$~GeV/$c^2$ and
with momentum above 0.8 (1.2)~GeV/$c$ for KLM (ECL) candidates.

To reconstruct $B^0 \rightarrow J/\psi K_L^0$,
we first use the KLM candidates. 
If none of the KLM candidates satisfy the selection
criteria below, we use the ECL candidates.
Thus, the two classes of $B^0 \rightarrow J/\psi K_L^0$ candidates are mutually exclusive.
In order to suppress the background, we calculate
a probability density function (pdf) for each of the following variables
and then form a product of the pdf's to obtain the combined likelihood:
the cms momentum of the $J/\psi$, $p^{\rm cms}_{J/\psi}$;
the angle between the $K_L^0$ candidate and the closest charged track having a
momentum larger than 0.7 GeV/$c$;
the Fox-Wolfram moment ratio $R_2$;
$\cos \theta_B$ where $\theta_B$ is the polar angle of the
reconstructed $B$ in the cms;
and the number of charged tracks with $p_{\rm t} > 0.1$ GeV/$c$, $|dr| < $ 2~cm,
and $|dz| <$ 4~cm.
In addition to these five variables, two other
variables are included conditionally
to reduce the background from $B^+ \rightarrow J/\psi K^{*+}$,
$K^{*+} \rightarrow K_L^0 \pi^+$ decays.
One such variable is
the cms momentum of the $J/\psi K_L^0\pi^+$ system, 
$p^{\rm cms}_B(J/\psi K_L^0 \pi)$, and
the other is the momentum of the additional pion.
We use
$p^{\rm cms}_B(J/\psi K_L^0 \pi)$ if
the invariant mass of the $K_L^0$ and a charged track, with
the nominal pion mass being assumed,
is above 0.85 GeV/$c^2$ and below 0.93 GeV/$c^2$, and
$p^{\rm cms}_B(J/\psi K_L^0 \pi) < 0.8$ GeV/$c$\cite{footnoteKLPB3}.
The pion momentum is used
if the addition of the
extra pion results in $0.2 < p^{\rm cms}_B(J/\psi K_L^0 \pi) < 0.45$ GeV/$c$
given the requirement above on
the $K_L^0\pi^+$ invariant mass.

The combined likelihood is calculated
for $J/\psi K_L^0$ (signal) and inclusive $B \rightarrow J/\psi$ decays (background)
using MC samples. 
Background events from misidentified (fake) $J/\psi$ mesons are not included in the likelihood construction
since their contribution is small.
We then form a likelihood ratio
${\cal L}_{J/\psi K_L^0} / ({\cal L}_{J/\psi K_L^0} + {\cal L}_{\rm bkg})$ 
that is used as a discriminant variable.
Figure~\ref{psikl_lh} shows the likelihood ratio distributions
for the data and Monte Carlo candidates.
\begin{figure}[!htb]
\begin{center}
\resizebox{0.48\textwidth}{!}{\includegraphics{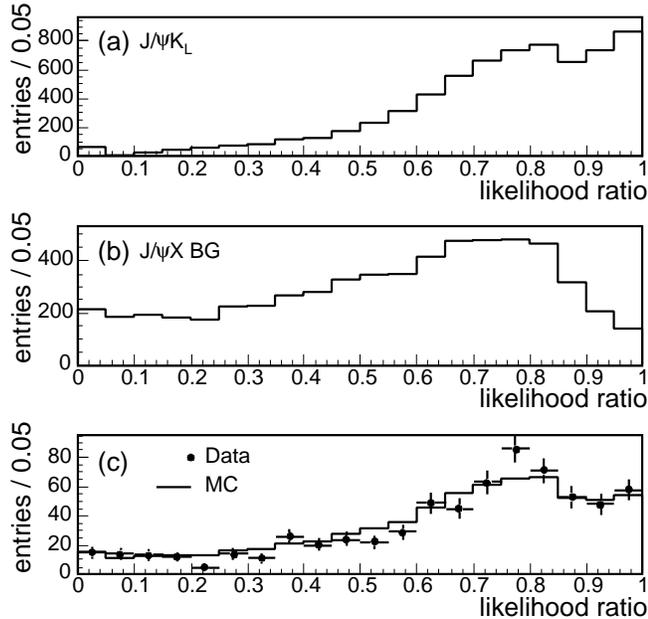}}
\vspace{1ex}
\caption{ The likelihood ratio distributions for
          (a) $J/\psi K_L^0$ Monte Carlo,
          (b) $J/\psi$ + X background Monte Carlo, and (c) the data
          in the signal region with Monte Carlo events overlaid.
}
\label{psikl_lh}
\end{center}
\end{figure}
We require that the likelihood ratio be larger than 0.4
to identify $J/\psi K_L^0$ candidates.

Figure~\ref{pbfit} shows the resultant
$p^{\rm cms}_B$ distribution for events that satisfy all the selection criteria. 
We define the signal region as
$0.2 \le p^{\rm cms}_B \le 0.45~(0.40)~{\rm GeV}/c$
for $K_L^0$ candidates identified with the KLM (ECL) criteria~\cite{footnoteMultiCand}, and
obtain 569 candidates,
out of which 397 are KLM candidates and 172 are ECL
candidates.

We extract the $J/\psi K_L^0$ signal yield 
by fitting the $p^{\rm cms}_B$ distribution of the data 
to a sum of four components:
 (1) signal; (2) background with $K_L^0$; (3) background without $K_L^0$; 
and (4) combinatorial $J/\psi$ mesons. 
The shapes of the first three components are determined from the $J/\psi$
inclusive MC 
and look-up tables are used in the fit.
The normalizations of these three components are treated as free 
parameters in the fit to minimize the effect of the aforementioned
uncertainty in
the $K^0_L$ detection efficiency in the MC simulation.
The combinatorial component is evaluated using events with $e\mu$-pairs
that satisfy the requirements
for $J/\psi$ reconstruction. 
The shape is modeled by a second-order polynomial.
An additional parameter of the fit is an offset in 
$p_B^{\rm cms}$, allowing the signal shape to shift with respect
to the background distribution. 

The result of the fit to 
the $p^{\rm cms}_B$ distribution
is shown in Fig.~\ref{pbfit}. 
By integrating each component obtained by the fit
in the signal region, 
we find a total of
($346.3 \pm 28.8$) $J/\psi K_L^0$ signal events, and a signal purity of
61\%.
\begin{figure}[!htb]
\begin{center}
\resizebox{0.48\textwidth}{!}{\includegraphics{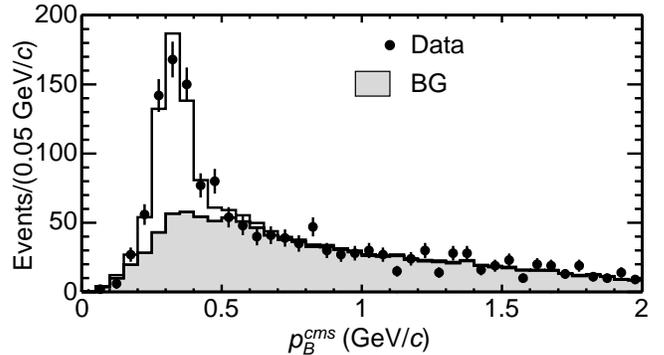}} 
\vspace{1ex}
\caption{The $p_B^{\rm cms}$ distribution of the $J/\psi K_L^0$ candidates.
         The solid line is a sum of the signal and background.
         The shaded histogram shows the background component only.}
\label{pbfit}
\end{center}
\end{figure}
The fitting procedure is applied to the KLM and ECL candidates separately.
The $\chi^2$ of the fit to KLM (ECL) candidates is 54.8 (31.4) for
$ndf$ = 35.
Table~\ref{fitresult} summarizes the fit results.

\begin{table}[!hbt]
\begin{center}
\caption {Results of the fit to the $p^{\rm cms}_B$ distribution of the $J/\psi K_L^0$
          candidates.
          Yields are for the signal $p^{\rm cms}_B$ region. }
\label{fitresult}
\begin{tabular}{lccc}
            &  KLM      & ECL   & Sum  \\
\hline
 Signal     
            & 244.8 $\pm$ 25.1
                & 101.5 $\pm$ 14.2      & 346.3 $\pm$ 28.8 \\
\hline
Background  & & & \\ 
 ~~with $K_L^0$    & 118.1 & 26.5 & 144.6 \\
 ~~without $K_L^0$        & 27.5 & 38.6 & 66.1 \\
 ~~Combinatorial        & 6.6 & 5.4 & 12.0 \\
\hline \hline
Total Yield &  397 & 172 & 569 \\
\end{tabular}
\end{center}
\end{table}

\subsection{\boldmath Control samples of flavor-specific $B$ decays} 
\label{subsec:control_samples}
The reconstruction of flavor-specific $B$ decays is also a
key ingredient in this analysis, since such decays are
used for various purposes:
evaluation of the performance of flavor tagging
(Section~\ref{sec:fbtgw});
extraction of the proper-time resolution function (Section~\ref{sec:resfunc});
$B$ lifetime measurements as a cross check (Section~\ref{subsubsec:b_lifetimes});
and the demonstration of null asymmetries
in non-$CP$ final states (Section~\ref{sec:contrl}).
Since a large number of events with high purity are
required for these purposes, we use
semileptonic decays and hadronic decays from $b \rightarrow c \overline{u} d$
transitions.

For the semileptonic decays,
we use the decay chains
$B^{0} \to D^{*-} \ell^{+} \nu$ and $D^{*-} \to \overline{D}{}^{0} \pi^-$,
where $\overline{D}{}^{0} \to K^+ \pi^-$,    
      $\overline{D}{}^{0} \to K^+ \pi^- \pi^0$
and   $\overline{D}{}^{0} \to K^+ \pi^+ \pi^-\pi^-$.
All tracks used must have associated SVD hits and radial
impact parameters $|dr|<0.2$~cm, except for the
slow pion from the $D^{*-}$ decay.
Candidate $\overline{D}{}^0$ decays are selected by requiring
the invariant mass to be consistent with
the nominal $\overline{D}{}^0$ mass. The invariant mass
requirement
depends on the $\overline{D}{}^0$ decay mode,
varying from $+9$ to $+23~{\rm MeV}/c^2$ above the $\overline{D}{}^0$ mass
and from $-9$ to $-37~{\rm MeV}/c^2$ below the $\overline{D}{}^0$ mass, respectively.
For the $D^{*-}$ reconstruction, we also impose a requirement on the
mass difference,
$M_{\rm diff}$,
between a $D^{*-}$ candidate and
the corresponding $\overline{D}{}^0$ candidate.
We require $M_{\rm diff}$ to be
within $0.8$ to $1.75$ ${\rm MeV}/c^2$ of the nominal mass difference,
depending on the $\overline{D}{}^0$ decay mode. 
In addition, we require the $D^{*-}$ cms momentum to be less than
2.6 GeV/$c$ to suppress $D^{*-}$ mesons from the continuum.
The $D^{*-}$ candidates are combined with $\mu^+$ or $e^+$
candidates having the opposite charge to the $D^{*-}$ candidate.
Lepton candidates must satisfy $1.4<p^{\rm cms}_{\ell}<2.4$ GeV/$c$,
where $p^{\rm cms}_{\ell}$ is the cms momentum of the lepton.
We exploit the massless character of the neutrino,
to calculate $M_{\rm miss}^2$, the effective missing mass squared in the cms defined by
$M_{\rm miss}^2 \equiv (E^{\rm cms}_B-E^{\rm cms}_{D^*\ell})^2-
|\vec{p}^{\rm ~cms}_B|^2-|\vec{p}^{\rm ~cms}_{D^*\ell}|^2$,
and a product of the momenta of the $B$ and the $D^*\ell$ system,
$C \equiv 2|\vec{p}^{\rm ~cms}_B|\,|\vec{p}^{\rm ~cms}_{D^*\ell}|$.
The cosine of the angle between 
$\vec{p}^{\rm ~cms}_B$ and $\vec{p}^{\rm ~cms}_{D^*\ell}$ is
given by $-M_{\rm miss}^2/C$.
In the $M_{\rm miss}^2$ versus $C$ plane, therefore, we select 
\dslnu\ candidates inside the triangle shown in Fig.~\ref{fig:cvsmm2}.
The triangle is defined by
$C \geq -\frac{1.45}{1.595}M_{\rm miss}^2$, 
$C \geq \frac{1.02}{1.1}M_{\rm miss}^2$ and
$C \leq -\frac{0.43}{2.695}(M_{\rm miss}^2+1.595)+1.45$.
\begin{figure}[htpb]
\begin{center}
\resizebox{0.48\textwidth}{!}{\includegraphics{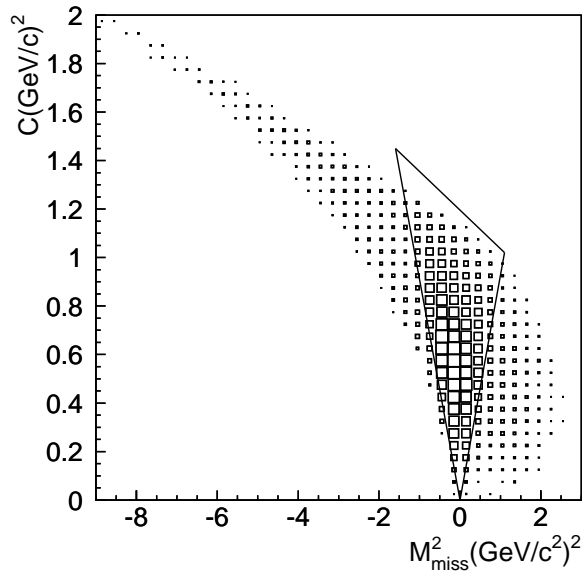}}
\vspace{1ex}
\caption{The distribution of candidate
         $B^{0} \to D^{*-} \ell^{+} \nu$ decays in the plane
         of $C$ versus $M_{\rm miss}^2$. The signal region (triangle) is also shown.}
\label{fig:cvsmm2}
\end{center}
\end{figure}

The vertex position of the $D^*\ell\nu$ candidate is obtained by first
fitting the vertex of the $D$ candidate from its daughter tracks,
and then fitting the lepton and $D$ candidate to obtain the 
$B$ vertex.
The slow pion track from the $D^*$ meson is also included
in the fit.
This is important since the
re-fitted helix parameters of the slow pion are used to
recalculate $M_{\rm diff}$ and improve its resolution.
The overall signal fraction in these control samples is estimated to be 
79.4\%. The backgrounds
consist of fake $D^*$ mesons (10.4\%), incorrect combinations of $D^*$ 
with leptons that do not show an angular correlation (4.1\%, called
``uncorrelated background''), continuum events (2.1\%) 
and $B^{\pm}$ decays (4\%).
The fraction of fake leptons in the uncorrelated background is
estimated with a Monte Carlo simulation
to be 0.5\% and is neglected.
We estimate the fake $D^*$ background by using
events in the $\overline{D}{}^0$ mass sideband as well as fake $D^*$ events 
that are reconstructed with wrong-charge slow pions.
We evaluate the uncorrelated background fraction by
counting signal events in a sample wherein
we flip the candidate lepton momentum vector artificially.
In this case the number of
uncorrelated background in the signal region
remains the same level while
the signal events are rejected~\cite{footnoteDSTARLNU}.
We apply the same event selection to off-resonance data 
( 2.3 fb$^{-1}$ ) and count the number of events in the
signal region after subtracting the fake $D^*$ background.
We estimate the continuum background fraction by scaling the
result with the integrated luminosity.
We fit the $C$ distribution with a range 
$-10 < C < 1.1$
to estimate the signal and $B\rightarrow D^{**}\ell \nu$ background
fractions. We use MC information to model the signal and 
$B\rightarrow D^{**}\ell \nu$ distribution.
The $C$ distributions and fractions of all the other backgrounds 
are obtained from the aforementioned special background
samples and are fixed in the fit.
We treat the background fraction uncertainties
as a source of systematic errors in the determination
of wrong tag fractions, which will be discussed
in Section~\ref{sec:fbtgw}.

For hadronic decays,
we use the decay modes $B^0 \to D^-\pi^+$, $D^{*-}\pi^+$ and
$D^{*-}\rho^+$,
where $D^- \to K^+\pi^-\pi^-$, $\rho^+ \to \pi^+\pi^0$.
We reconstruct $D^{*-}$ candidates in the same modes that were used for the
$D^{*-}\ell^+ \overline\nu$ mode.
For $D$ and $D^{*-}$ candidates we apply mode-dependent requirements on
the reconstructed $D$ mass (ranging from $\pm 30$ to $\pm 60$ MeV/$c^2$)
and $M_{\rm diff}$ (ranging from $\pm 3$ to $\pm 12$ MeV/$c^2$),
in a similar way as for $D^{*-}\ell^+ \overline\nu$ mode.
We select $\rho^+$ candidates by requiring 
the $\pi^+\pi^0$ invariant mass
to be within 150 MeV/c$^2$ of the nominal $\rho^+$ mass.
In order to suppress continuum background, we impose mode dependent cuts on 
$R_2$ (upper cut values ranging from 0.5 to 1.0) and 
$\cos\theta_{\rm thr}$ (upper cut values ranging from 0.92 to 1.0).
The cut values are chosen to maximize the figure of merit 
$S/\sqrt{S+B}$ for each mode, where $S$ and $B$ are numbers of signal
and background, respectively. 
We select $B^0$ candidates by requiring
5.27 $< M_{\rm bc} <$ 5.29 GeV/$c^2$ and $|\Delta E| <$ 50 MeV. 
Background contributions are estimated by fitting the $M_{\rm bc}$ and $\Delta E$
distributions in the same way as we do for $CP$ modes. 
Figure~\ref{fig:controlsample_mbc} shows the $M_{\rm bc}$ distribution 
for the three control sample modes 
combined. The overall signal purity is $\sim$82\%.
We study background components with a MC sample that includes
both $B\overline{B}$ and continuum events.
We find no significant peaking background in the signal region.
Therefore we use the ARGUS function to model
the background in the $M_{\rm bc}$ distribution.
A possible deviation due to combinatorial backgrounds
may introduce ambiguities in the signal fraction
determination and the background $\Delta t$ model parameters
of the control sample, resulting in small uncertainties
in the final determination of the $CP$ asymmetry.
This is considered as a source of systematic uncertainties.
\begin{figure}[htpb]
\begin{center}
\resizebox{0.48\textwidth}{!}{\includegraphics{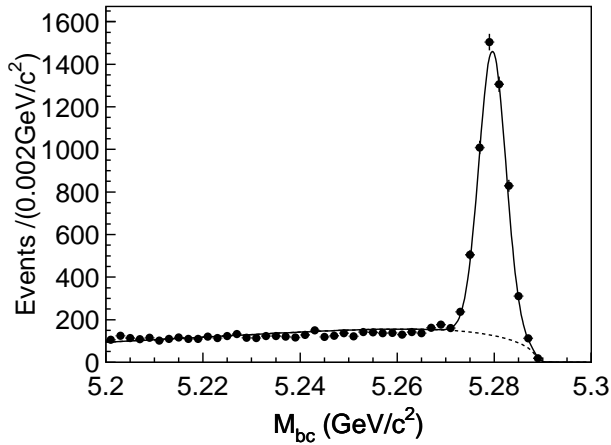}}
\vspace{1ex}
\caption{ $M_{\rm bc}$ distribution for all hadronic decay modes
 ($B^0 \to D^-\pi^+$,  $D^{*-}\pi^+$, and $D^{*-}\rho^+$)
  with $|\Delta E| <$ 50 MeV.
  Fit curves for background (dashed) and signal plus background events (solid)
  are superimposed. }
\label{fig:controlsample_mbc}
\end{center}
\end{figure}                                        

The numbers of candidates and their purities for flavor-specific samples
are summarized in Table~\ref{Nev_flv}
\begin{table}[htb]
\caption{Summary of the numbers of flavor-specific decay candidates 
 and their estimated purities. } 
\medskip
\label{Nev_flv} 
\begin{tabular}{lcc}
 Decay mode & Candidates & Purity \\
\hline
 $B^{0} \to D^{*-} \ell^{+} \nu$  & 16101  & 79.4 \% \\
\hline
 $B^{0} \to D^{-} \pi^{+} $       &  2241  & 85.5 \% \\
 $B^{0} \to D^{*-} \pi^{+} $      &  2126  & 87.2 \% \\
 $B^{0} \to D^{*-} \rho^{+} $     &  1620  & 72.2 \% \\
\end{tabular}
\end{table}

\section{Flavor tagging}
\label{sec:flavor_tagging}

After the exclusive reconstruction of a neutral $B$ meson decaying
into a final state, $f_{CP}$, all the remaining particles should belong to
the final state, $f_{\rm tag}$, of the decay of the other $B$ meson.
To observe time-dependent $CP$ violation,  we need to ascertain whether
$f_{\rm tag}$ is from a $B^0$ or 
$\bzb$.
This determination is called ``flavor tagging."
The simplest and most reliable method for flavor tagging uses
the charge of high-momentum leptons in
semileptonic decays, i.e.
$B^0 \rightarrow X \ell^+ \nu$ and 
$\bzb \rightarrow X \ell^- \overline \nu$.
The charges of final-state kaons can also be used 
since the decays
$B^0 \rightarrow K^+ X$ (with
$\overline b \rightarrow \overline c \rightarrow \overline s$)
and
$\bzb \rightarrow K^- X$ (with 
$b \rightarrow c \rightarrow s$)
dominate. 
In addition to these two leading discriminants, our
algorithm includes other categories of tracks
whose charges depend on the $b$ quark's flavor: 
lower momentum leptons from $c \rightarrow s \ell^+ \nu $;
$\Lambda$ baryons from the cascade decay $b \rightarrow c \rightarrow s$;
high-momentum pions that originate from decays like
$B^0 \rightarrow D^{(*)-}(\pi^+,\rho^+,a_1^+,$ etc.);
and slow pions from $D^{*-} \rightarrow \overline D{}^0 \pi^-$.
All these inputs are combined, taking their correlations
into account, in a way that maximizes the flavor tagging performance.
The performance is characterized  by two parameters, $\epsilon$ and $w$.
The parameter $\epsilon$ is the raw tagging efficiency,
while $w$ is the probability that the flavor tagging is wrong
(wrong tag fraction).
A non-zero value
of $w$ results in a dilution of the true asymmetry.
For example, if the true numbers of reconstructed $B^0$ and 
$\bzb$ are $n_{B^0}$ and 
$n_{\bzb}$, the corresponding asymmetry is 
${\cal A} = (n_{B^0}-n_{\bzb})/(n_{B^0}+n_{\bzb})$.
With realistic flavor tagging, the observed numbers are
$N_{B^0} = \epsilon( (1-w)n_{B^0} + w n_{\bzb} )$ for $B^0$,
$N_{\bzb} = \epsilon( (1-w)n_{\bzb} + w n_{B^0} )$ 
for $\bzb$, and
the observed asymmetry becomes $(1-2w){\cal A}$. 
Since the statistical error of the measured asymmetry is
proportional to $\epsilon^{-1/2}$,
the number of events required to observe the asymmetry 
for a certain statistical
significance is proportional to 
$\epsilon_{\rm eff} = \epsilon (1-2w)^2$, which is called
the ``effective efficiency.'' 
Note that an imperfect knowledge of $w$ shifts the central value
of the measurement and thus represents a potential source
of systematic error. 

In light of the above, our tagging algorithm has been
designed to maximize $\epsilon_{\rm eff}$.  Moreover,
since $w$ directly affects the central value of our result,
we have developed an approach wherein it can be determined
from the data.   

In our approach, we use two parameters,
$q$ and $r$, to represent the tagging information.
The first, $q$, corresponds to the sign of
the $b$ quark charge where $q = +1$ for $\overline b$ and hence
$B^0$, and $q = -1$ for $b$ and $\bzb$.
The parameter $r$ is an event-by-event flavor-tagging dilution factor 
that ranges from $r=0$ for no flavor
discrimination to $r=1$ for unambiguous flavor assignment.
The values of $q$ and $r$ are determined for each event from
a look-up table.
Each entry of the table is prepared using a large statistics MC sample
and is given by
\begin{equation}
q\cdot r \equiv \frac{N(B^0) - N(\bzb)}{N(B^0) + N(\bzb)},
\label{tag-q}
\end{equation} 
where $N(B^0)$ and $N(\bzb)$ are the numbers of 
$B^0$ and $\bzb$ in the MC sample.

In this analysis, we sort flavor-tagged events into six bins in $r$.
For each $r$ bin, we empirically determine $w$ 
directly from data  
by using control samples, as described in Section~\ref{sec:fbtgw} below.

\subsection{Flavor Tagging Method }
\label{sec:fbtg_method}

Flavor tagging proceeds in two stages.  In the first stage, 
the flavor tagging information ($q$ and $r$) provided by each track in the
event is calculated.   In the second, the track-level results
are combined to determine event-level values for $q$ and $r$.

Tracks are sorted into four categories, namely those that resemble
leptons, kaons, $\Lambda$ baryons, and slow pions.
For each category, we consider several tagging discriminants, 
such as track momentum and particle identification information.
The value of $q$ and $r$ for each track is assigned based
on MC-generated look-up tables that take the tagging discriminants
as input.  

In the second stage,
the results from the four track categories are combined to determine
the values of $q$ and $r$ for each event.
Again a look-up table is prepared to provide $q\cdot r$. 

Figure \ref{total} shows a schematic diagram of 
the flavor tagging method.
The event-level parameter $r$ should satisfy
$r \simeq 1-2w$ where we measure $w$ from control samples.
Using this MC-determined dilution factor $r$ as a measure
of the tagging quality is a straightforward and
powerful way of taking into account correlations among various discriminants.
Using two stages, we keep the look-up tables small enough to provide sufficient MC statistics
for each entry.
In the following, we provide additional details about each stage of the
flavor tagging.

\begin{figure}[htpb]
\begin{center}
\resizebox{0.48\textwidth}{!}{\includegraphics{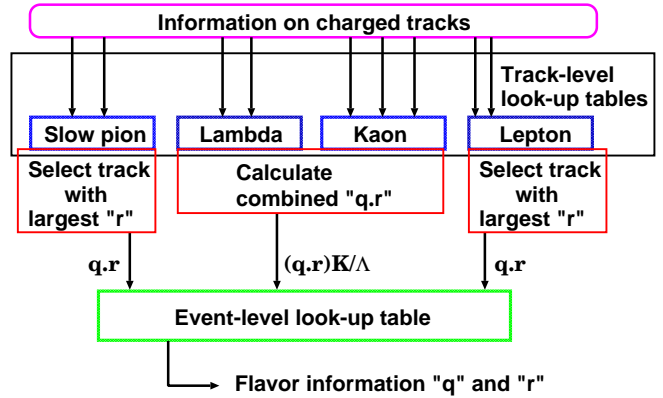}}
\vspace{1ex}
\caption{A schematic diagram of the two-stage flavor tagging.
See the text on the definition of the parameters ``$q$" and ``$r$".}
\label{total}
\end{center}
\end{figure}

\subsubsection{Track-level flavor tagging}

We select tracks that do not belong to $B_{CP}$ and
that satisfy $|dr| < 2$~cm and $|dz| < 10$~cm. 
Tracks that are part of a $K_S^0$ candidate are not used.
Each selected tag-side track is examined and assigned to one of the four track categories.
Tracks in the lepton category are subdivided into categories for electron-like
and muon-like tracks.
If the cms momentum, $p^{\rm cms}_\ell$, of a track is larger than 0.4 GeV/$c$ and 
the ratio of its electron and kaon likelihoods 
is larger than 0.8, the track is assigned to the electron-like category. 
If a track has $p^{\rm cms}_\ell$ larger than 0.8 GeV/$c$ and the ratio of its muon and kaon 
likelihoods is larger than 0.95, it is passed to the muon-like category.
The likelihood is calculated by combining the ACC, TOF, $dE/dx$, and
ECL or KLM information.
In the lepton category, leptons from semileptonic $B$ decays yield the largest
effective efficiency. Leptons from $B\rightarrow D$ cascade decays and
high-momentum pions from $B^0 \rightarrow D^{(*)-}\pi^+X$ also make 
a small contribution to this category.
We choose the following six discriminants:
  the track charge;
  the magnitude of the momentum in the cms, $p^{\rm cms}_\ell$;
  the polar angle in the laboratory frame, $\theta_{\rm lab}$;
  the recoil mass, $M_{\rm recoil}$, 
      calculated using all the tag side tracks except the lepton candidate;
  the magnitude of the missing momentum in the cms, $P^{\rm cms}_{\rm miss}$;
  and the  lepton-ID quality value.

The track charge directly provides
the $b$-flavor $q$.
The lepton-ID quality distinguishes leptons from pions.
Its performance is reinforced by
variables $p^{\rm cms}_\ell$ and $\theta_{\rm lab}$, 
which have distributions that are different for leptons and pions.  
The variables $p^{\rm cms}_\ell$, $M_{\rm recoil}$ and $P^{\rm cms}_{\rm miss}$
discriminate between high momentum leptons from semileptonic $B$
decays and intermediate momentum leptons from 
$B \rightarrow D$ cascade decays where the $D$
decays semi-leptonically.

If a track cannot be positively identified as a kaon
and its momentum is less than 0.25 GeV/$c$,
it is assigned to the slow-pion category,
since low-momentum pions often come from 
charged $D^* \rightarrow D\pi$ decays.  
Here the discriminant variables are:
the track charge;
the momentum and polar angle 
in the laboratory frame,
$p_{\rm lab}$ and $\theta_{\rm lab}$; 
the ratio of the electron to $\pi$ probability from $dE/dx$,
and $\cos\alpha_{\rm thr}$, the
cosine of the angle between the slow pion candidate and 
the thrust axis of the tag-side particles in the cms.
The main background in this category comes from other ({\it i.e.} non-$D^*$ daughter) low momentum pions,
electrons from photon conversions and $\pi^0$ Dalitz
decays.
To separate slow pions from those electrons we use $dE/dx$ for this class.
Since the direction of the slow pion from a $D^*$ decay 
is approximately parallel to
the $D^*$ direction,
it is also almost parallel to the thrust axis.
The variables $\cos\alpha_{\rm thr}$,
$p_{\rm lab}$ and $\theta_{\rm lab}$, thus, help
to identify the slow pions originating from $D^*$ decays.

If a track forms a $\Lambda$ candidate with another track, it is assigned 
to the $\Lambda$ category.  In that category the discriminant variables are:
the flavor ($\Lambda$ or $\overline \Lambda$);
the invariant mass of the reconstructed $\Lambda$ candidate;
the angle difference between the $\Lambda$ momentum vector and 
 the direction of the $\Lambda$ vertex point from the nominal IP;
the mismatch in the $z$ direction of the two tracks at the $\Lambda$ vertex point;
and the proton-ID quality value.

If a track does not fall in any of the categories described above, and
is not positively 
identified as a proton, it is classified as a kaon.  
The kaon category  is subdivided into two parts,
one for events with 
$K_S^0$ decays, and the other for events without $K_S^0$'s.
Separate treatment is necessary as
events with $K_S^0$ have a larger wrong tag fraction
because of their additional strange quark content.
We use
the track charge, $p^{\rm cms}$, $\theta_{\rm lab}$ and the probability ratio
of kaon to pion
as the tagging discriminants.
The charge of kaons is the most important discriminant.
The other three variables help separate kaons from pions.

Although the discriminating power of high-momentum pions is weaker
than that of charged kaons, they do provide some tagging information
and are therefore included in the kaon category. 
Approximately half of the pions with $p^{\rm cms}>1.0~{\rm GeV}/c$ are
included in the kaon category, while the other half falls into 
the lepton class, mostly in the muon-like category.

As is described in Section~\ref{sec:fbtgw},
we have developed a method wherein the wrong tag fractions
in our flavor tagging method are evaluated from data.
Thus possible discrepancies between data and MC
in the distributions of discriminant variables do not
affect our $\sin 2\phi_1$ measurement, although they might
result in the degradation in the effective efficiency.
Nevertheless, we have made detailed comparisons between
data and MC, which are described elsewhere~\cite{FBTG},
and obtain consistent distributions.

\subsubsection{Event-level flavor tagging}

The event-level flavor tagging combines the results from
each of the track categories to determine
an overall $q$ and $r$. 
For the lepton and slow-pion track categories, 
we take the $b$-flavor assignment 
from the track with the highest $r$-value in each category. 
For the kaon and $\Lambda$ categories, 
a combined $b$-flavor output is calculated as the product
of likelihood values for all tracks:
\begin{equation}
(q \cdot r)_{K/\Lambda} 
              = \frac{\prod_i[1+(q \cdot r)_i] - \prod_i[1-(q \cdot r)_i]}
                     {\prod_i[1+(q \cdot r)_i] + \prod_i[1-(q \cdot r)_i]}.
\end{equation}
where the subscript $i$ runs over all tracks in the kaon and $\Lambda$ categories. 
The product likelihood is designed to 
use the information from the sum of the strangeness,
which provides better flavor-tagging performance than simply
choosing the best 
candidate.

Using the three aforementioned track-level $q \cdot r$ values,
the event-level $q$ and $r$ values are obtained from a look-up table
that is prepared with
a MC sample that is independent of the sample used to obtain
$q\cdot r$ values in the track categories.
The probability that we can assign a non-zero value for $r$ 
is 99.6\% in MC; i.e. almost all the reconstructed candidates
can be used to extract $\sin 2 \phi_1$.

We specify the following six regions :
$0<r\leq 0.25$, $0.25 < r \leq 0.5$,  $0.5 < r \leq 0.625$,
$0.625 < r \leq 0.75$, $0.75 < r \leq 0.875$
and $0.875 < r \leq 1$.
For each region we obtain the wrong tag fraction,
$w_l$, where $l$ is the region ID ($l = 1,...,~6$),
using hadronic and semileptonic control samples,
which is described in the next section.
In this way, the analysis is insulated from
systematic differences between the MC simulation and the data due
to imperfections in the modeling of the detector response,
decay branching
fractions, and fragmentation in our MC simulation.

As a validation, we compare the distribution
of $q\cdot r$ in the $D^*\ell\nu$ control sample
with the MC expectation. As shown in
Fig.~\ref{fig:mdlh_data_vs_mc}, 
the data and MC are in good agreement.
\begin{figure}[htpb]
\begin{center}
\resizebox{0.48\textwidth}{!}{\includegraphics{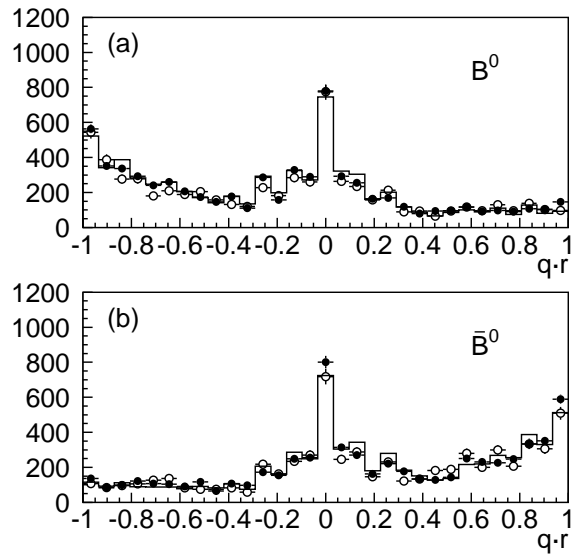}}
\vspace{1ex}
\caption{The $q\cdot r$ distribution for
(a) $B^0$ and (b) $\bzb$ candidates. 
In each figure the closed (open) points with errors show the 
$B^0 \to D^{*-}\ell^+\nu$ (hadronic $B^0$ decays) data 
with the background subtracted, while the histogram is the MC prediction.}
\label{fig:mdlh_data_vs_mc}
\end{center}
\end{figure}

\subsection{Flavor Tagging Performance}
                            \label{sec:fbtgw}
The flavor tagging performance is evaluated
by replacing the $CP$-eigenstate side of the event with a flavor-specific decay
and tagging the $b$-flavor for the other side using the method described above.
We use the semileptonic decay
$B \rightarrow D^{*}\ell\nu$
and hadronic modes $B^0 \rightarrow D^{(*)-}\pi^+$, and $D^{*-}\rho^+$
for this purpose.
The overall efficiency of our flavor tagging
is 99.7\% which is consistent with the MC expectation.

Since we know the flavors of both $B$ mesons in this case,
we can observe
the time evolution of neutral $B$-meson pairs with
opposite flavor (OF) or same flavor (SF), 
which is given by:
\begin{eqnarray}
{\cal P}_{\rm OF}(\Delta t) &\propto& 1 + (1-2w)\cos(\Delta m_d \Delta t),
  \nonumber\\   
{\cal P}_{\rm SF}(\Delta t) &\propto& 1 - (1-2w)\cos(\Delta m_d \Delta t),
  \nonumber   
\end{eqnarray}
and the OF-SF asymmetry,
$$ A_{\rm mix} \equiv { {\cal P}_{\rm OF} - {\cal P}_{\rm SF} \over 
               {\cal P}_{\rm OF} + {\cal P}_{\rm SF} } 
           = (1 - 2w)\cos(\Delta m_d \Delta t),      $$
where $w$ is the wrong tag fraction.
We thus obtain the value of $w$ directly from the data 
by measuring the amplitude of the OF-SF asymmetry.

We obtain the wrong tag fraction
by fitting the $\Delta t$ distribution of the SF and OF
events, with
$\Delta m_d$ fixed at the world average value of
0.472 ps$^{-1}$\cite{PDG2000}.
The procedure to form the probability density function
(pdf) for the fit is
quite similar to that adopted for the 
maximum likelihood analysis of $CP$ eigenstates,  
which is described in the next section.

The resolution function for signal events, which models how the true distribution is
smeared by the finite vertex resolution, is constructed by fitting the proper-time
distributions without discriminating between the OF and SF events
and with the lifetime fixed to the world average value.
In the fit we use the background fraction estimated for each region of $r$,
and the proper-time distribution for background obtained
using events outside the signal region. 
For hadronic modes the sideband regions in $M_{\rm bc}$ and $\Delta E$
are used. For semileptonic decays
the upper sideband in $M_{\rm diff}$ is used
for the fake $D^*$ backgrounds. Uncorrelated backgrounds are
modeled with the events that are found in the signal region 
after inverting the momentum of the lepton.
Semileptonic decays $D^*X\ell\nu$ are treated as signal events
since they approximately obey the same OF-SF asymmetry.

Figure~\ref{fbtg_asym} shows the measured OF-SF asymmetries as a function of
$\Delta t$ for tagged $D^{*\mp}\ell^\pm\nu$ 
events for the six regions of $r$. 
The curves in the figure are obtained by the fit.
The background is not subtracted in the plots.
\begin{figure}[htpb]
\begin{center}
\resizebox{0.48\textwidth}{!}{\includegraphics{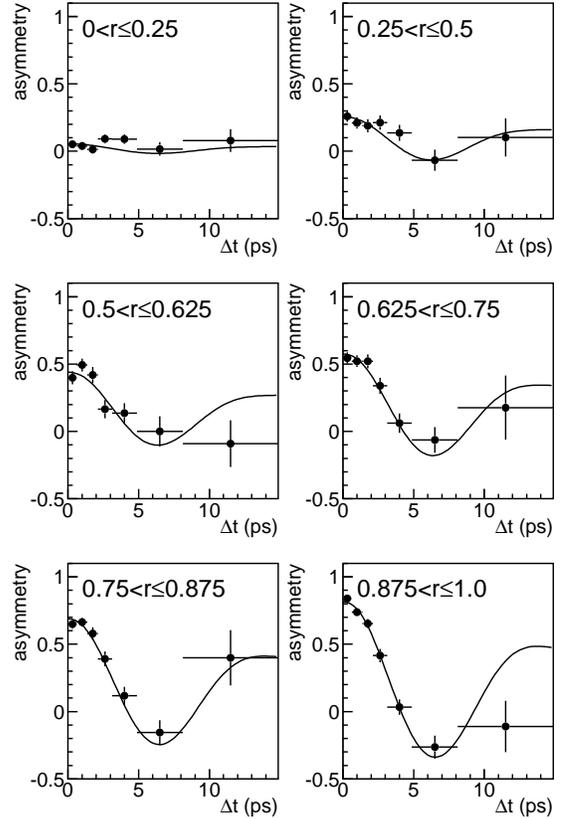}}
\vspace{1ex}
\caption{Measured asymmetries between the OF events and the SF events
         (OF-SF asymmetries)
         for six regions of $r$ obtained for the control sample
         $B^0 \rightarrow D^*\ell\nu$. The definition of the OF and SF events
         is given in the text. The background is not subtracted in the asymmetry
         plots.
         Fit curves are also shown.}
\label {fbtg_asym}
\end{center}
\end{figure}                                        
For hadronic modes the fits to OF and SF events
are similar to those in the semileptonic case.

We also fit signal MC samples to examine the difference
between the generated and reconstructed values.
We apply small corrections to $w_l$, that correspond to the difference. 
For hadronic modes 
the corrections range from 0.003 to 0.03 depending on the $r$ region. 
For semileptonic decays 
the difference is consistent with zero within statistical errors,
and we apply no correction

To combine the results from semileptonic and hadronic decays,
we calculate the weighted average and its error.
We conservatively treat
the difference between the weighted average 
and each measurement as an additional systematic
error, and add this difference in quadrature with the error.
The systematic errors for the semileptonic mode are dominated by the
uncertainties on the background fractions
and are comparable to the statistical errors.
As explained in Section~\ref{subsec:control_samples},
the background estimation relies little on MC information since we
use control samples whenever possible. 
One important exception is the $C$ distribution of the $D^{**}\ell\nu$
background; we use several $D^{**}$ components and add them
with fixed fractions using MC. Since these fractions are poorly
known experimentally, we conservatively assume that 
each component dominates the $D^{**}\ell\nu$ distribution
and repeat the fit procedure to obtain the systematic error.
For the hadronic modes, the main contribution to the systematic error
comes from the
uncertainty of the fit bias obtained from the MC simulation, but 
the statistical errors dominate.
The event fractions and wrong tag fractions
are summarized in Table~\ref{tag_sum}.
\begin{table}
\caption{
The event fractions ($\epsilon_l$)
and wrong tag fractions ($w_l$)
for each $r$ interval. The errors include both statistical
and systematic uncertainties. The event fractions are obtained from 
the $J/\psi K_S^0$ simulation.}
\label{tag_sum}  
\begin{tabular}{lccccc}
$l$&$r$ & $\epsilon_l$ & $w_l(D^*\ell\nu)$& $w_l$(hadronic)& $w_l$(combined)\\
\hline
1&$0.000-0.250$ & $0.405$ & $0.463^{+0.011}_{-0.011}$ & 
  $0.469^{+0.015}_{-0.016}$ & $0.465^{+0.010}_{-0.009}$ \\
2&$0.250-0.500$ & $0.149$ & $0.351^{+0.019}_{-0.017}$ & 
  $0.352^{+0.026}_{-0.026}$ & $0.352^{+0.015}_{-0.014}$ \\
3&$0.500-0.625$ & $0.081$ & $0.254^{+0.021}_{-0.020}$ & 
  $0.219^{+0.031}_{-0.030}$ & $0.243^{+0.021}_{-0.030}$\\
4&$0.625-0.750$ & $0.099$ & $0.169^{+0.019}_{-0.018}$ & 
  $0.192^{+0.028}_{-0.027}$ & $0.176^{+0.022}_{-0.017}$\\
5&$0.750-0.875$ & $0.123$ & $0.107^{+0.015}_{-0.015}$ &
  $0.127^{+0.032}_{-0.031}$ & $0.110^{+0.022}_{-0.014}$\\
6&$0.875-1.000$ & $0.140$ & $0.041^{+0.012}_{-0.011}$ &
  $0.041^{+0.024}_{-0.023}$ & $0.041^{+0.011}_{-0.010}$
\end{tabular}
\end{table}
The total effective efficiency obtained by summing over the $r$ regions
is calculated to be
\[
\epsilon_{\rm eff} =
\sum_l \epsilon_l(1-2w_l)^2 = (27.0 \pm 0.8({\rm stat}) ^{+0.6}_{-0.9}({\rm syst}))\%,
\] 
where $\epsilon_l$ is the event fraction in each of the six regions.

Our simulation indicates that events with high-momentum leptons dominate
the highest $r$ region and provide the cleanest tagging
information. Events with charged kaons have lower $r$, but are more numerous, 
and thus provide the largest contribution to 
the effective tagging efficiency.
The effective efficiency using each category alone is examined with MC.
We obtain 
13\% for the lepton category, 19\% for the kaon and lambda categories combined, and
4\% for the slow pion category. Note that 
the sum of these values exceeds $\epsilon_{\rm eff}$ (29.6\% in MC)
since an event contains tracks in different categories.

We check for possible biases in the flavor tagging by measuring the 
effective tagging efficiency for the $B^0$ and $\bzb$ control samples
separately, and for the $q=+1$ and $-1$ samples separately.
We find no statistically significant difference.

\section{Maximum likelihood Fit}
\label{sec:maximum_likelihood_fit}

We determine 
$\sin 2\phi_1$ 
by performing an
unbinned maximum-likelihood fit of a $CP$ violating
probability density function (pdf) to the observed $\Delta t$ 
distributions. These pdf's come from the theoretical distributions
diluted and smeared by the detector response.
For modes other than $J/\psi K^{*0}$
the pdf expected for the signal is
\begin{equation}  \label{pdf_sig_1}
{\cal P}_{\rm sig}(\Delta t,q,w_l,\xi_f) =
\frac{ e^{-|\Delta t|/\tau_{B^0}} }{2\tau_{B^0}}
\PCP(\Delta t,q,w_l,\xi_f),
\end{equation}
where
\begin{equation}   \label{pdf_sig_2}
\PCP(\Delta t,q,w_l,\xi_f) = 1- \xi_f q (1-2w_l)\sin 2\phi_1\sin(\Delta m_d\Delta t).
\end{equation}
In order to take into account the effect of finite vertex resolution on
the $\Delta t$ distribution,
this pdf is convolved with a resolution function, $R_{\rm sig}(\Delta t)$.
Our vertex reconstruction method is explained in Section~\ref{sec:vtx}.
The parametrization and extraction of $R_{\rm sig}(\Delta t)$
are described in Section~\ref{sec:resfunc}.
We also incorporate the effect of background
that dilutes the significance of $CP$ violation in the time distribution of Eq.~(\ref{pdf_sig_1}).
The $\Delta t$ distribution for background events, 
$P_{\rm bkg}(\Delta t)$, is constructed in a similar way to 
the signal distribution and is described in detail in Section~\ref{sec:rbkg}.

As a result, we adopt the following $\Delta t$ distribution function
for each event:
\begin{eqnarray}
&P(\Delta t_i;&\sin 2\phi_1)  \nonumber \\
&& = f_{\rm sig}{\displaystyle
\int}{\cal P}_{\rm sig}(\Delta t^\prime,
q,w_l,\xi_f) R_{\rm sig}(\Delta t_i-\Delta t^\prime) d\Delta t^\prime \nonumber \\
&& +(1-f_{\rm sig})P_{\rm bkg}(\Delta t_i),
\end{eqnarray}
where $f_{\rm sig}$ is the probability that the event is signal,
being calculated for each candidate
from $p_B^{\rm cms}$ for $J/\psi K_L^0$ and
a combination of $\Delta E$ and $M_{\rm bc}$ for other modes.
The only free parameter is $\sin 2\phi_1$, which is
determined by maximizing
the likelihood function
\begin{equation}
 L=\prod_{i} P(\Delta t_i; \sin 2\phi_1),
 \label{eq:likeli}
\end{equation}
where the product is over all candidates. 
We perform a blind analysis: 
The fitting algorithms were developed and finalized 
without using the flavor information $q$.

In the following we explain the details
of $R_{\rm sig}(\Delta t)$, $P_{\rm bkg}(\Delta t)$ and $f_{\rm sig}$ in turn.
The likelihood for $J/\psi K^{*0}(\rightarrow K_S^0 \pi^0)$ candidates
is described separately in Section~\ref{sec:LH_jpsikstar}.

\def\rsig{R_{\rm sig}}
\def\micron{$\mu$m}
\def\delz{\Delta z}
\def\dt{\Delta t}
\def\dzb{\delz_B}
\def\dtb{\dt_B}
\def\dtp{\dt'}
\def\DTb{$\dtb$}
\def\Dzb{$\dzb$}
\def\Dz{$\delz$}
\def\dtrec{\dt_{\rm rec}}
\def\dtgen{\dt_{\rm gen}}
\def\sigdt{\sigma_{\dt}}
\def\sigmisdt{\sigma^{\rm tail}_{\dt}}
\def\sigdtbg{\sigma_{\dt}^{\rm bkg}}
\def\sigmisdtbg{\sigma^{\rm tail,bkg}_{\dt}}
\def\sigdz{\sigma_{\delz}}
\def\sigmisdz{\sigma^{\rm tail}_{\delz}}
\def\sigdzbg{\sigma_{\delz}^{\rm bkg}}
\def\sigmisdzbg{\sigma^{\rm tail,bkg}_{\delz}}
\def\acp{\alpha_{CP}}
\def\atag{\alpha_{\rm tag}}
\def\sigzcp{\sigma_{z}^{CP}}
\def\sigztag{\sigma_{z}^{\rm tag}}
\def\sigtzcp{\tilde{\sigma}_{z}^{CP}}
\def\sigtztag{\tilde{\sigma}_{z}^{\rm tag}}
\def\sigk{\sigma_{K}}
\def\sigmisk{\sigma^{\rm tail}_{K}}
\def\smis{S_{\rm tail}}
\def\scharm{S_{\rm charm}}
\def\sdet{S_{\rm det}}
\def\sdetbg{S_{\rm det}^{\rm bkg}}
\def\sdata{S_{\rm det}}
\def\scmis{S^{\rm tail}_{\rm charm}}
\def\sdmis{S^{\rm tail}_{\rm det}}
\def\sdmisbg{S^{\rm tail}_{\rm bkg}}
\def\mudz{\mu_{\delz}}
\def\mumisdz{\mu^{\rm tail}_{\delz}}
\def\muz{\mu_0}
\def\mumisz{\mu^{\rm tail}_0}
\def\mudt{\mu_{\dt}}
\def\mumisdt{\mu^{\rm tail}_{\dt}}
\def\mudtbg{\mu_{\dt}^{\rm bkg}}
\def\mumisdtbg{\mu^{\rm tail,bkg}_{\dt}}
\def\amu{\alpha_\mu}
\def\amismu{\alpha^{\rm tail}_{\mu}}
\def\fmis{f_{\rm tail}}
\def\fmisbg{f_{\rm tail,bkg}}
\def\bg{\beta\gamma}
\def\bgu{(\beta\gamma)_\Upsilon}
\def\chisqndf{{\chi}^2/n}
\def\chisq{\chi^2}
\def\Chisqndf{$\chisqndf$}
\def\Chisq{$\chisq$}
\def\dstb{\overline{D}^{*}}
\def\jpsi{J/\psi}
\def\bdstlnu{B\to\dstb\ell^+\nu}
\def\bpsik{B\to\jpsi K}
\def\dz{D^0}
\def\Dz{$\dz$}
\def\piz{\pi^0}
\def\pip{\pi^+}
\def\pim{\pi^-}
\def\kz{K^0}
\def\kp{K^+}
\def\km{K^-}
\def\ks{K_S^0}
\def\bm{{B^-}}
\def\Bzb{$\bzb$}
\def\Bm{$\bm$}
\def\etal{{\it et al.}}

\subsection{Vertex Reconstruction}
\label{sec:vtx}

The decay vertices for the $CP$ side that include a $J/\psi$ candidate
are reconstructed using leptons 
from the $J/\psi$ and a constraint on the $B$ decay point. 
The $B$ decay point is constrained by the measured
profile of the interaction point (IP profile)
convolved with the finite $B$ flight length in
the plane perpendicular to the $z$ axis (the $r$-$\phi$ plane).
The IP profile is represented by 
a three-dimensional Gaussian distribution.
The standard deviation of each Gaussian is determined
using pre-selected $B\overline{B}$ candidates
on a run-by-run basis, while the mean is evaluated
in finer subdivisions. 
The typical size 
of the IP profile is 100~$\mu$m in $x$, 5~$\mu$m in 
$y$ and 3~mm in $z$. Since the size in the $y$ direction
is too small to be measured from the vertex distribution,
it is taken from special measurements by the KEKB accelerator group.
For leptons, we require that there are sufficient 
SVD hits associated with a CDC track by a Kalman filter technique;
i.e. both $z$ and $r$-$\phi$ hits in at least 
one layer and at least one additional layer with a $z$ hit.
In order to remove events with mis-reconstructed tracks,
we require that the reduced $\chi^2$ 
(\Chisqndf, $n={}$number of degrees of freedom)
of the vertex be less than 20. 
The vertex reconstruction efficiency is 
measured to be 95\% with 
$B^{\pm} \rightarrow J/\psi K^{\pm}$ and $B^0 \rightarrow J/\psi
K^{*0}(\rightarrow K^{\pm}\pi^{\mp})$ 
events. This is consistent with
the expectation from the SVD acceptance and cluster matching efficiency.
The resolution estimated by MC is typically 75$\mu$m (rms).

For $B \rightarrow \eta_c K_S^0$ candidates, the method is basically
the same as for $J/\psi$, replacing $\ell^+ \ell^-$
with $K^+ K^-$ in case of $\eta_c \rightarrow K^+ K^- \pi^0$
and with $K^{\pm} \pi^{\mp}$ for
$\eta_c \rightarrow K_S^0 K^{\pm} \pi^{\mp}$.
Although the resolution in these cases is
worse than for candidates with a $J/\psi$ vertex, it is still better than
the tag-side vertex resolution.

The algorithm for tag-side vertex reconstruction is chosen
to minimize the effect of long-lived particles, secondary vertices from
charmed hadrons and a small fraction of poorly reconstructed tracks.
From all the charged tracks except those used to reconstruct
the $CP$ side, we 
select tracks that have associated SVD hits in the same way 
as for the $CP$ side. We also require that the
impact parameter with respect to the $CP$-side vertex be less than 0.5~mm in the $r$-$\phi$ 
plane, less than 1.8~mm in $z$, and
the vertex error in $z$ be less than 0.5~mm.
Tracks are removed if they form a $K_S^0$ candidate satisfying 
$|M_{K_S^0} - M_{\pi^+\pi^-}| < 15~{\rm MeV}/c^2$.
Tracks satisfying these criteria
are used to reconstruct the tag-side vertex where
the IP constraint is also applied.
If the reduced $\chi^2$ of the vertex is good, we accept
this vertex. Otherwise we remove the track
that gives the largest contribution to the $\chi^2$ and
repeat the vertex reconstruction. 
If the track to be removed is a lepton
with $p^{\rm cms}_\ell > 1.1~{\rm GeV}/c$,
however, we keep the lepton and
remove the track with the second worst $\chi^2$.
This trimming procedure is repeated
until we obtain a good reduced $\chi^2$.
The reconstruction efficiency was measured to be 93\% for
$B^{\pm} \rightarrow J/\psi K^{\pm}$ and $B^0 \rightarrow J/\psi
K^{*0}(\rightarrow K^+\pi^-)$ candidates, 
consistent with the MC expectation (91\%).
The resolution estimated from the simulation is typically 140$\mu$m (rms).

\subsection{Signal Resolution Function}
\label{sec:resfunc}
The resolution function $\rsig(\dt -\dtp)$ is parametrized 
by the sum of two Gaussians:
\begin{eqnarray}
\rsig(\dt -\dtp)&=&
(1-\fmis)G(\dt - \dtp;\mudt,\sigdt)
\nonumber \\
&+&\fmis G(\dt-\dtp;\mumisdt,\sigmisdt),
\end{eqnarray}
where $G(x;\mu,\sigma)$ is a Gaussian distribution in $x$ with mean $\mu$ and rms $\sigma$.
The parameter $\fmis$ describes the fraction of the tail of the resolution function,
and $\sigdt$, $\sigmisdt$, $\mudt$ and $\mumisdt$ are the proper-time
difference resolutions and the mean value shifts of the proper-time
difference for the main part and the tail of the resolution function,
respectively.  The value of $\fmis$ is determined to be $0.03\pm0.02$ from the lifetime
analysis of hadronic samples using the same resolution function.  

The proper-time difference resolutions $\sigdt$ and $\sigmisdt$ are 
calculated on an event-by-event basis taking into account 
the error $\sigk$
in the kinematic approximation
$\dt\approx\frac{\delz}{c\bg},\ \bg=\frac{p_z(\Upsilon)}{m(\Upsilon)}$:
\begin{eqnarray*}
\sigdt&=&\sqrt{(\frac{\sigdz}{c\bg})^2+\sigk^2},\\
\sigmisdt&=&\sqrt{(\frac{\sigmisdz}{c\bg})^2+(\sigmisk)^2}.\\
\end{eqnarray*}
We measure $\sigk=0.287\pm0.004$~ps 
and $\sigmisk=0.32\pm0.19$~ps using the MC simulation.  These parameters are 
independent of the detector performance.

The parameters $\sigdz$ and $\sigmisdz$ are calculated from the 
event-by-event vertex errors
of the two $B$ mesons, $\sigzcp$ and $\sigztag$, which
are computed from the track helix errors in the vertex fit.
We use 
\begin{eqnarray*}
\sigdz^2&=&\sdet^2(\sigzcp)^2+(\sdet^2+\scharm^2)(\sigztag)^2,\\
(\sigmisdz)^2&=&(\sdmis)^2(\sigzcp)^2+\{(\sdmis)^2+(\scmis)^2\}(\sigztag)^2,
\end{eqnarray*}
where $\scharm$ and $\scmis$ are scaling factors to account for the degradation  
of the vertex resolution on the tag-side due to contamination from 
charm daughters, and $\sdet$ and $\sdmis$ are global scaling factors that account 
for systematic uncertainties in the vertex errors $\sigzcp$ and $\sigztag$. 
We determine $\scharm=0.58 \pm 0.01$ 
and $\scmis=2.16 \pm 0.10$ using the MC simulation.
The values of $\sdet$ and $\sdmis$ are measured from the data as they
depend on the detector performance.
We determine $\sdet$ using a $\dz\to\kp\pim$ control sample.
The production point of the \Dz\ is obtained from the primary tracks in the same 
hemisphere as the \Dz\ candidate using the IP constraint. The distance between the \Dz\ 
decay vertex and the production vertex in the $z$ direction is fit with the same resolution 
function and the known \Dz\ lifetime to obtain $\sdet$.
We measure $\sdet=0.88\pm0.01$ from the data
and $\sdet=1.05\pm0.01$ from a MC simulation of the \Dz\ sample.
Finding $\sdet=1.035\pm0.003$ for a $\bpsik$ MC sample, we use 
$\sdet=(0.88\pm0.01)\times(1.035\pm0.003)/(1.05\pm0.01)=0.86\pm0.01$ for the data.
We determine $\sdmis$ to be 
$3.51\pm0.88$ from the lifetime analysis of the flavor-specific hadronic samples.

For $B \rightarrow \eta_c K_S^0$ decays, we introduce an additional scale factor
to account for the difference between the $\eta_c$ and $J/\psi$ decays.
Using the MC, we determine the additional scale factor for $\eta_c$ to be
$1.07\pm0.02$ for both $\eta_c \rightarrow K^+ K^- \pi^0$ and
$\eta_c \rightarrow K_S^0 K^{\pm} \pi^{\mp}$.

A small fraction of events have a large reduced $\chi^2$.
We have found that the vertex error computed from track helix errors in the 
vertex fit underestimates the vertex resolution and the vertex with larger 
\Chisq\ has worse resolution.
In order to take into account this effect, we introduce effective vertex resolutions 
$\sigtzcp$ and $\sigtztag$ when \Chisqndf\ is greater than 3:
\begin{eqnarray}
(\sigtzcp)^2&=&[1+\acp\{(\chisqndf)_{CP}-3\}](\sigzcp)^2,
\nonumber \\
(\sigtztag)^2&=&[1+\atag\{(\chisqndf)_{\rm tag}-3\}](\sigztag)^2,
\label{eq:sigt}
\end{eqnarray}
where $(\chisqndf)_{CP}$ and $(\chisqndf)_{\rm tag}$ are the reduced \Chisq\ of the vertex fits 
for the $CP$ and tagging $B$ decay vertices, respectively.
The coefficients $\acp = 1.02 \pm 0.03$ and $\atag = 1.64 \pm 0.05$ are determined by
a MC study of $B^0 \rightarrow J/\psi K_S^0$.

As mentioned above, the offsets $\mudt$ and $\mumisdt$ originate from 
the mean shifts of the $\delz$ measurements $\mudz$ and $\mumisdz$, respectively:
\begin{eqnarray*}
\mudt&=&\frac{\mudz}{c\bg}, \\
\mumisdt&=&\frac{\mumisdz}{c\bg}.
\end{eqnarray*}
The mean value shifts, $\mudz$ and $\mumisdz$, are caused by 
contamination from charm daughters in the vertex reconstruction on the tag-side 
and are correlated with $\sigztag$:
\begin{eqnarray*}
\mudz(\sigztag)&=&\muz+\amu\sigztag, \\
\mumisdz(\sigztag)&=&\mumisz+\amismu\sigztag.
\end{eqnarray*}
The values for $\muz$ and $\mumisz$ are determined 
from hadronic samples to be
$\muz=(-21.4\pm3.7)~\mu$m and $\mumisz=(151\pm128)$~\micron,
while $\amu$ and $\amismu$ are
derived from MC simulation where we obtain
$\amu=-0.10\pm 0.01$
and $\amismu=-1.42\pm0.17$.

Figure~\ref{fig:dt-fit} (a) shows the $\dtrec-\dtgen$ distribution for 
the MC $J/\psi K_S^0(\pi^+ \pi^-)$ candidates along with the resolution function, 
where $\dtrec$ and $\dtgen$ are the reconstructed and true proper-time 
differences, respectively.
The resolution function is obtained by summing event-by-event
resolution functions.  
The distribution is well-represented by the resolution function.
The average resolution function obtained from 
the $J/\psi K_S^0(\pi^+\pi^-)$ data is shown in Fig.~\ref{fig:dt-fit} (b),
which is represented by the sum of two Gaussian distributions
with the following parameters:
$\mu_{\rm main}$= $-0.24$ ps, $\mu_{\rm tail}$= 0.18 ps, $\sigma_{\rm main}$= 1.49 ps,
$\sigma_{\rm tail}$= 3.85 ps.

\begin{figure}[tbhp]
\begin{center}
\resizebox{0.48\textwidth}{!}{\includegraphics{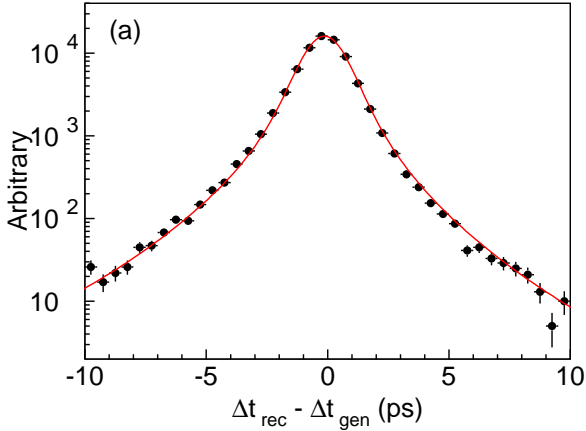}}
\resizebox{0.48\textwidth}{!}{\includegraphics{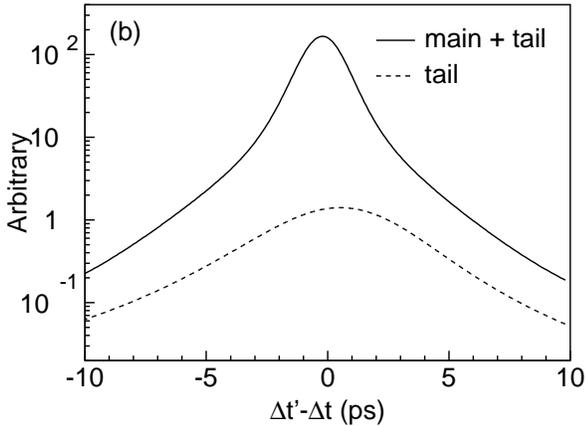}}
\vspace{1ex}
\caption{(a) The $\dtrec-\dtgen$ distributions for the MC $J/\psi K_S^0(\pi^+ \pi^-)$
candidates
  with the resolution function and (b)
 the average resolution function from $J/\psi K_S^0(\pi^+\pi^-)$ data.
 The vertical scales are arbitrary.
}
\label{fig:dt-fit}
\end{center}
\end{figure}

\subsection{Background Shape}  \label{sec:rbkg}
The background likelihood function is defined in a similar
way to the signal function,
\begin{equation}
P_{\rm bkg}(\Delta t) = \int^{+\infty}_{-\infty} {\cal P}_{\rm bkg}(\Delta t^\prime)
\cdot R_{\rm bkg}(\Delta t-\Delta t^\prime)  d\Delta t^\prime.
\end{equation}
Although $R_{\rm bkg}(\Delta t-\Delta t^\prime)$ is treated as a resolution
function for the background, it does not need to be an exact model of
the vertex resolution. It is more important that $P_{\rm bkg}$ 
represents the proper-time distribution of the whole background sample
with sufficient precision.
 
The pdf for background events is expressed as
\begin{equation}
{\cal P}_{\rm bkg}(\Delta t)
 =
 f_{\tau} \frac{ e^{-|\Delta t|/\tau_{\rm bkg}} }{2\tau_{\rm bkg}}
 + (1-f_{\tau})\delta(\Delta t^\prime),
\end{equation}
where 
$f_{\tau}$ is the fraction of the background component with 
an effective lifetime of $\tau_{\rm bkg}$, and $\delta$ is the Dirac delta function.  
We assume no asymmetry in the background $\Delta t$ distribution.
We find $f_{\tau}$ to be small using background-dominated regions in
the $\Delta E$ versus $M_{\rm bc}$ plane of $J/\psi K_S^0$ and $J/\psi K^+$
candidates. We thus use
${\cal P}_{\rm bkg}(\Delta t)=\delta(\Delta t)$ for all $f_{CP}$ modes except
for $J/\psi K_L^0$.

The background in the $J/\psi K_L^0$
mode is dominated by $B \to J/\psi X$ decays, including
\textit{CP} eigenstates that have to be treated differently from
non-\textit{CP} states.  The ${\cal P}_{\rm bkg}$ for the $J/\psi K_L^0$ mode is
determined by a MC simulation study
separately for each background component: 
$J/\psi K^{*0}(K_L^0\pi^0)$, $\xi_f=-1$ \textit{CP} modes ($J/\psi K_S^0$),
$\xi_f=+1$ \textit{CP} modes ($\psi(2S) K_L^0$, $\chi_{c1} K_L^0$ and 
$J/\psi \pi^0$), the other $B^0$, $B^\pm$ decays and combinatorial
background.  For the \textit{CP}-mode backgrounds, we use the signal pdf
given in Eq.~(\ref{pdf_sig_2}) with the appropriate $\xi_f$ values.  
For the $J/\psi K^{*0}(K_L^0\pi^0)$
mode, which is a mixture of $\xi_f=-1$ (about $81\%$) and $\xi_f=+1$ (about $
19\%$) states~\cite{Itoh}, we use a net \textit{CP} eigenvalue of
$\xi_{\psi K^{*0}}=-0.62\pm0.11$.

Accordingly, we obtain the background pdf for $J/\psi K_L^0$
\begin{eqnarray}
&{\cal P}_{\rm bkg}&(\Delta t) \equiv   \left.
   \frac{ e^{-|\Delta t|/\tau_{B^0}} }{2\tau_{B^0}} \times
   \right\{ f_{{\rm bg}B^0}  
\nonumber \\
&& + f_{{\rm bg}\psi K^{*0}}\PCP(\Delta t,q,w_l,\xi_{\psi K^{*0}})
\nonumber \\
&&   + f_{{\rm bg}CP_{\rm odd}}\PCP(\Delta t,q,w_l,-1) 
\\
&& + f_{{\rm bg}CP_{\rm even}}\PCP(\Delta t,q,w_l,+1) \left\}
   + f_{{\rm bg}B^\pm}\frac{ e^{-|\Delta t|/\tau_{{\rm bg}B^\pm}} }{2\tau_{{\rm bg}B^\pm}} \right.
\nonumber \\
&& + f_{\rm bgCmb}\left\{ f_{\tau {\rm Cmb}} 
      \frac{ e^{-|\Delta t|/\tau_{\rm bgCmb}} }{2\tau_{\rm bgCmb}} 
      + (1-f_{\tau {\rm Cmb}}) \delta( \Delta t ) \right\},
\nonumber
\end{eqnarray}
where $f_{{\rm bg}\psi K^{*0}}$, $f_{{\rm bg}CP_{\rm odd}}$, $f_{{\rm bg}CP_{\rm even}}$, 
$f_{{\rm bg}B^0}$, $f_{{\rm bg}B^\pm}$, and $f_{\rm bgCmb}$
are the fractions of background components from $J/\psi K^{*0}(K_L^0\pi^0)$, 
$\xi_f=-1$ \textit{CP}-modes, $\xi_f=+1$ \textit{CP}-modes,
the remaining $B^0$, 
$B^\pm$, and combinatorial respectively
($f_{{\rm bg}\psi K^{*0}} + f_{{\rm bg}CP_{\rm odd}} + 
f_{{\rm bg}CP_{\rm even}} + f_{{\rm bg}B^0} + f_{{\rm bg}B^\pm} + f_{\rm bgCmb} = 1$).
The fraction of each background component is a function of $p^{\rm cms}_B$. 
The fraction, $f_{\rm bgCmb}$, is calculated as described in Section~\ref{sec:psikl_rec}.
The combinatorial background includes
a prompt component with the fraction of 
$(1-f_{\tau {\rm Cmb}})$ where $f_{\tau {\rm Cmb}}$~=~0.26 $\pm$ 0.08.
The lifetime distribution of the combinatorial background is obtained
from $e$-$\mu$ combinations with invariant masses in the $J/\psi$ region
that satisfy our selection criteria.
The effective lifetime is determined to be 
$\tau_{\rm bgCmb} = 1.03\pm0.12$~ps,
also from the $e$-$\mu$ control sample.
A MC study shows that the effective lifetime for background from $B^{\pm}$, 
$\tau_{{\rm bg}B^\pm}$, is shorter than the $B^{\pm}$ lifetime due to the contamination of 
charged tracks from $f_{CP}$ (mostly $\pi^{\pm}$ from $J/\psi K^{*\pm}(K_L^0 \pi^{\pm})$) 
into the tag-side vertex. The value of
$\tau_{{\rm bg}B^\pm}$ is determined from the MC simulation to be ($1.49\pm0.04$)~ps.
The same MC study shows that the effective lifetime for $B^0$ backgrounds is 
consistent with the nominal $B^0$ lifetime.
Thus we use the nominal $B^0$ lifetime in our fit.

For the $J/\psi X$ background in the $J/\psi K_L^0$ mode, we use
the signal resolution function $R_{\rm sig}$ 
to model the background since both the \textit{CP}- and tag-side vertices are reconstructed
with similar combinations of tracks for these backgrounds.

For the combinatorial background, we use 
\begin{eqnarray}        \label{eqn:rbkg}
&R_{\rm bkg}(\dtp -\dt)& =
(1-\fmisbg)G(\dtp-\dt; \mudtbg, \sigdtbg)
\nonumber \\
&& + \fmisbg G(\dtp-\dt; \mumisdtbg, \sigmisdtbg),
\end{eqnarray}
where $\fmisbg$, $\mudtbg$, and $\mumisdtbg$ are constants 
determined from data.
The resolutions $\sigdtbg$ and $\sigmisdtbg$ are calculated on an event-by-event basis as
\begin{eqnarray*}
\sigdtbg &=& \frac{\sigdzbg}{c\bg}, \\
\sigmisdtbg & = & \frac{\sigmisdzbg}{c\bg}, \\
(\sigdzbg)^2 &= & (\sdetbg)^2 [ (\sigtzcp)^2+(\sigtztag)^2 ], \\
(\sigmisdzbg)^2 &=& (\sdmisbg)^2  [ (\sigtzcp)^2+(\sigtztag)^2 ],
\end{eqnarray*}
where $\sigtzcp$ and $\sigtztag$ are calculated as shown in Eq.~(\ref{eq:sigt}).

We use different values of $\mudtbg$ and $\mumisdtbg$ for the finite lifetime component and the 
zero-lifetime component, since they come from different 
types of events.
The background shape parameters for all modes except $J/\psi K_L^0$ are obtained 
from events in the background-dominated regions of $\Delta E$ versus $M_{\rm bc}$.
For $J/\psi K_L^0$, we use events with $e$-$\mu$ pairs 
to determine the properties of fake $J/\psi$ candidates, as discussed in Section~\ref{sec:psikl_rec}.
The parameters used in the fit are summarized in Table~\ref{bgshape}.

\begin{table}[htbp]
\caption{Background shape parameters for the combinatorial background.}
\begin{center}
\begin{tabular}{ccc} 
parameters & $c\overline{c} K_S^0(K^{*0})$ & $J/\psi K_L^0$ \\ \hline
$\sdetbg$ & $1.08\pm0.06$ & $0.88\pm0.07$ \\
$\sdmisbg$& $3.31\pm0.28$ & $3.94\pm1.14$ \\
$\fmisbg$ & $0.14\pm0.04$ & $0.05\pm0.02$ \\
lifetime component & & \\
$\mudtbg$(ps)    & N/A & $-0.33\pm0.10$ \\
$\mumisdtbg$(ps) & N/A & $1.86\pm1.28$ \\
prompt component & & \\
$\mudtbg$(ps)    & $-0.05\pm0.04$ & $-0.22\pm0.15$ \\
$\mumisdtbg$(ps) & $-0.12\pm0.26$ & $-5.00\pm1.10$ \\ 
\end{tabular}
\end{center}
\label{bgshape}
\end{table}

\subsection{Signal probability}
  \label{sec:signalprob}
The signal probability, $f_{\rm sig}$, is calculated as a function of $\Delta E$ and
$M_{\rm bc}$ for each event.
It is given by
\begin{eqnarray}
  f_{\rm sig}(\Delta E, M_{\rm bc}) & = & \frac{F_{\rm SIG}(\Delta E, M_{\rm bc})}
   {F_{\rm BG}(\Delta E, M_{\rm bc})+
  F_{\rm SIG}(\Delta E, M_{\rm bc})},
\end{eqnarray}
where $F_{\rm SIG}(\Delta E, M_{\rm bc})$ is the signal function
and $F_{\rm BG}(\Delta E, M_{\rm bc})$ is
the background function.

In the case of \bjpsiks\, 
each distribution of
$\Delta E$ and $M_{\rm bc}$ is well modeled
by a Gaussian function.
For the background, we use a linear function for $\Delta E$ and
the ARGUS parametrization\cite{argus_func} for $M_{\rm bc}$:
\begin{eqnarray}
&F_{\rm SIG}&(\Delta E,M_{\rm bc}) 
\nonumber \\
&& =  a \cdot
  G(\Delta E; \mu_{\Delta E}, \sigma_{\Delta E}) \cdot
  G(M_{\rm bc}; \mu_{M_{\rm bc}}, \sigma_{M_{\rm bc}}), 
\nonumber \\
&F_{\rm BG}&(\Delta E,M_{\rm bc}) 
\nonumber \\
&& =  b \cdot (1 + c \cdot \Delta E) \cdot
      M_{\rm bc}\sqrt{1-(M_{\rm bc}/E_{\rm beam})^2}
\nonumber \\
&& \times   \exp(n \cdot [1-(M_{\rm bc}/E_{\rm beam})^2])
\end{eqnarray}
where $a$ and $b$ are normalization factors consistent with the overall
signal-to-background ratios obtained from the fit to the $M_{\rm bc}$
distribution in the $\Delta E$ signal region.
The values $\sigma_{M_{\rm bc}}$, $\mu_{M_{\Delta E}}$, 
$\sigma_{M_{\Delta E}}$, $\mu_{M_{\rm bc}}$, $c$ and $n$ are 
determined from a fit to the data.

The $\Delta E$ and $M_{\rm bc}$ distributions for 
\bpsipks ($\psi(2S) \to \ell^+\ell^-$), \bpsipks (\psipjpsipp), and 
$B^0 \to \eta_{c} K_S^0$ ($\eta_c \to K_S^0 K^{\pm}\pi^{\mp}$), 
are determined using the same procedure as that for
\bjpsiks.  For \bchiciks, the $\Delta E$ and $M_{\rm bc}$ distributions are
determined from MC simulation because 
the data sample for this mode is too small to estimate the parameters reliably.

The treatment for modes that include 
$\pi^0$ mesons such as \bjpsiks($K_S^0 \to \pi^0 \pi^0$) and
$B^0 \to \eta_{c} K_S^0$($\eta_{c} \to K^+ K^- \pi^0$), 
is different. While the fit function for the $M_{\rm bc}$ distribution remains the same,
the $\Delta E$ distributions are better represented by
the Crystal Ball function\cite{cb-line}:
\[  
 \left\{ \begin{array}{@{\,}ll}
  \frac{1}{\rm A} \exp(-\frac{(\Delta E - \mu_{\Delta E})^2}
   {2\sigma^2_{\Delta E}})
   &{\rm  for}~\Delta E > \mu_{\Delta E} - \alpha\sigma_{\Delta E} \\
          \frac{1}{\rm A} \frac{\exp\left({-\alpha^2/2}\right)}
           {\left[1-\frac{(\Delta E-\mu_{\Delta E})\alpha}
             {\sigma_{\Delta E}\alpha}-\frac{\alpha^2}{n}\right]^n}
           & {\rm for}~\Delta E < \mu_{\Delta E} - \alpha\sigma_{\Delta E}.
         \end{array}
         \right.
\]
All the parameters for these fits were determined from MC simulation because the number of events
for these modes in data is too small.
The integrated background fractions in the signal region
are listed in Table~\ref{bkg_frac}.

For the $B^0 \rightarrow J/\psi K_L^0$ fit, we define the signal
probability as a function of $p^{\rm cms}_B$, as described in 
Section \ref{sec:psikl_rec}. 

\begin{table}[htb]
\caption{Summary of the numbers of candidates and background fraction
in the signal region for each mode.
The values are obtained for events that have successful vertex reconstruction
and flavor tagging.
}
\medskip
\label{bkg_frac} 
\begin{tabular}{lcc}
 Decay mode      & Events  & bkg. fraction  \\ \hline
$B^0 \rightarrow J/\psi K_S^0, ~K_S^0 \rightarrow \pi^+\pi^- $
& 387 & 0.038 $\pm$ 0.010 \\
$B^0 \rightarrow J/\psi K_S^0, ~K_S^0 \rightarrow \pi^0\pi^0 $
&  57 & 0.272 $\pm$ 0.054 \\
$B^0 \rightarrow \psi (2S)K_S^0,~\psi (2S) \rightarrow \ell^+\ell^- $
&  33 & 0.038 $\pm$ 0.028 \\
$B^0 \rightarrow \psi (2S)K_S^0,~\psi (2S) \rightarrow J/\psi \pi^+\pi^-  $
&  32 & 0.078 $\pm$ 0.027 \\
$B^0 \rightarrow \chi_{c1}K_S^0,~\chi_{c1} \rightarrow J/\psi \gamma$
&  17 & 0.144 $\pm$ 0.056 \\
$B^0 \rightarrow \eta_c K_S^0, ~\eta_c \rightarrow K_S^0 K^\pm \pi^\mp$
&  35 & 0.242 $\pm$ 0.045 \\
$B^0 \rightarrow \eta_c K_S^0, ~\eta_c \rightarrow K^+ K^- \pi^0$
&  17 & 0.560 $\pm$ 0.164 \\
$B^0 \rightarrow J/\psi K_L^0$
& 523 & 0.379 $\pm$ 0.048 \\
$B^0 \rightarrow J/\psi K^{*0}, ~K^{*0} \rightarrow K_S^0 \pi^0 $
&  36 & 0.163 $\pm$ 0.054 \\
\end{tabular}
\end{table}

\subsection{\boldmath Likelihood for $J/\psi K^{*0} (\rightarrow K_S^0 \pi^0$)}
\label{sec:LH_jpsikstar}

For the $B^0 \rightarrow J/\psi K^{*0}$ fit, the signal pdf we use is: 
\begin{eqnarray}
&{\cal P}_{\rm sig}&(\Delta t,\theta_{\rm tr},q,w_l,\xi_f) =   
 \frac{ e^{-|\Delta t|/\tau_{B^0}} }{2\tau_{B^0}} \times
\nonumber \\
 & & [(1-f_{\rm odd})\frac{3}{8}(1+\cos^2\theta_{\rm tr})\PCP(\Delta t,q,w_l,+1) 
\nonumber \\
 & & +f_{\rm odd} \frac{3}{4}(1-\cos^2\theta_{\rm tr})\PCP(\Delta t,q,w_l,-1)],
\end{eqnarray}
where $f_{\rm odd}$ is the fraction of $\xi_f=-1$ decays in the 
\bjpsikstar($K^{*0} \to K_S^0 \pi^0$) mode determined from a full
angular analysis to be $0.19 \pm 0.04 \pm 0.04$~\cite{Itoh}. Here $\theta_{\rm tr}$ 
is defined in the transversity basis\cite{TRANSVERSITY}
as the angle between the positive $J/\psi$ decay lepton direction and the axis
normal to the $K^{*0}$ decay plane in the $J/\psi$ rest frame. 
$\PCP(\Delta t,q,w_l,\xi_f)$ is defined in Eq.~(\ref{pdf_sig_2}).
The signal resolution function is identical to that used for the other modes.
For the background shape, we also use Eq.~(\ref{eqn:rbkg})
for $R_{\rm bkg}$ except for the $J/\psi X$ background where we use $R_{\rm sig}$ 
in the same way as for the $J/\psi K_L^0$ fit.

We use the following background pdf:
\begin{eqnarray}
{\cal P}_{\rm bkg}(\Delta t) &\equiv& 
  f_{\rm bgFA} \frac{ e^{-|\Delta t|/\tau_{B_{\rm FA}}} }{2\tau_{B_{\rm FA}}} 
+ f_{\rm bgNR} \frac{ e^{-|\Delta t|/\tau_{B_{\rm NR}}} }{2\tau_{B_{\rm NR}}} 
\nonumber \\
&+& f_{\rm bgCmb}\delta( \Delta t ),
\end{eqnarray}
where $f_{\rm bgFA}$, $f_{\rm bgNR}$, and $f_{\rm bgCmb}$ are the fractions of
background components from feed-across from other $J/\psi K^{*}$ modes,
non-resonant $B^0 \to J/\psi K_S^0 \pi^0$ decays and combinatorial
background.  The fractions of feed-across and non-resonant decays are
determined from the MC simulation and from $K^*$ mass sideband events, respectively,
and are functions of $M_{\rm bc}$.  The fraction of combinatorial background is
determined in the same way as for the $K_S^0$ modes. The effective lifetimes of
the feed-across and non-resonant decay backgrounds, $\tau_{B_{\rm FA}}$ and
$\tau_{B_{\rm NR}}$, are fixed to the $B^0$ lifetime in the fit.

Finally, the determination of $f_{\rm sig}$
follows the method for other modes that include $\pi^0$ mesons.
The $\Delta E$ distribution is modeled by a Crystal Ball function.
We consider contributions from the
feed-across from other \bjpsikstar\ modes as well as from the non-resonant 
$B^0 \to J/\psi K_S^0 \pi^0$ mode, which make a peak in the signal region.  
These background fractions are determined from the MC simulation and \ksta\ 
mass sideband data, respectively.

\section{Fit results}
\label{sec:fit_results}
The likelihood fit is applied to the 1137 candidates
where the vertex reconstruction and flavor tagging have been successful. 
We obtain
\begin{eqnarray*}
 \sin 2\phi_1 & = & 0.99\pm0.14({\rm stat}).
\end{eqnarray*}
The observed $CP$ violation is large.
Figure~\ref{fig:ptd-chaks} shows the $\Delta t$ distributions 
together with the results from the fit.
Indeed, the broken $CP$ symmetry is visually apparent from
the difference 
between the number of events for
$q\xi_f = +1$ and $q\xi_f = -1$
at each $\Delta t$ bin,
despite the dilution from the vertex resolution, 
background events and incorrect flavor tagging.

We examined the value of $\sin 2 \phi_1$ in various sub-samples.
Applying the likelihood fit to $(c\overline{c})K_S^0~(\xi_f = -1)$ and
$J/\psi K_L^0~(\xi_f = +1)$ separately, we obtain
$\sin 2\phi_1 = 0.84\pm0.17({\rm stat})$ and $1.31\pm0.23({\rm stat})$, respectively.
Figure~\ref{fig:loglikeli} shows the log-likelihood values as a function
of $\sin 2\phi_1$ for $CP$-odd, $CP$-even, and all decay modes.
A more detailed breakdown along with separate results for $q=+1$ and $-1$ is
given in Table~\ref{cp_result}. We find no systematic trends 
beyond statistical fluctuations.
\begin{table}[htb]
\caption{ Summary of $\sin 2\phi_1$ fit results. 
          Only statistical errors are shown.}
\medskip
\label{cp_result} 
\begin{tabular}{lrl}
Sample & Events & $\sin 2\phi_1$\\
\hline
$f_{\rm tag}=B^0$ ($q=+1$) & 560 &  $0.84\pm 0.21$\\
$f_{\rm tag}=\overline{B}{}^0$ ($q=-1$) & 577 & $1.11\pm 0.17$\\
\hline
$(c\overline{c})K_S^0$          & 578 & $0.84\pm 0.17$\\
$J/\psi K_S^0(\pi^+\pi^-)$ & 387 & $0.81\pm 0.20$\\
$(c\overline{c})K_S^0$ except $J/\psi K_S^0(\pi^+\pi^-)$ & 191 & $1.00 \pm 0.40$\\
$J/\psi K_L^0$  & 523 & $1.31\pm 0.23$\\
$J/\psi K^{*0}(K_S^0\pi^0)$\cite{footnoteKstar} & 36 & $0.97 \pm 1.40$\\
\hline 
All & 1137 & $0.99\pm 0.14$
\end{tabular}
\end{table}

Figure~\ref{fig:asym} shows the 
asymmetry, $\sin 2\phi_1 \cdot \sin (\Delta m_d \Delta t)$, obtained 
in each $\Delta t$ bin for (a) all modes,  (b) $CP$-odd modes and (c) $CP$-even modes.
The unbinned maximum likelihood fit is performed 
separately for events
in each $\Delta t$ bin.  The value of $\sin 2\phi_1$ and
its error are multiplied by the average value of $\sin (\Delta m_d \Delta t)$
in each $\Delta t$ bin of the plot. 
The points are plotted at the average $\Delta t$ of each bin.

\begin{figure}[htbp]
 \begin{center}
  \resizebox{0.48\textwidth}{!}{\includegraphics{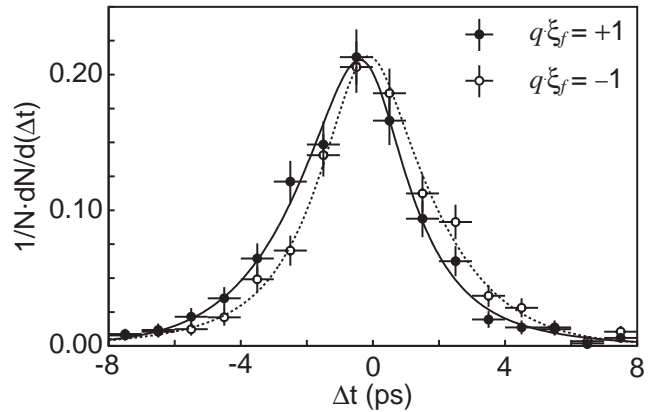}}
  \vspace{1ex}
  \caption{
  $\Delta t$ distributions 
for the events with $q\xi_f = +1$ (solid
points) and $q\xi_f = -1$ (open points). The 
results of the global fit (with  $\sin 2\phi_1 = 0.99$)
are shown as solid and dashed curves, respectively.
   }
  \label{fig:ptd-chaks}
 \end{center}
\end{figure}
\begin{figure}[htbp]
 \begin{center}
  \resizebox{0.48\textwidth}{!}{
  \includegraphics{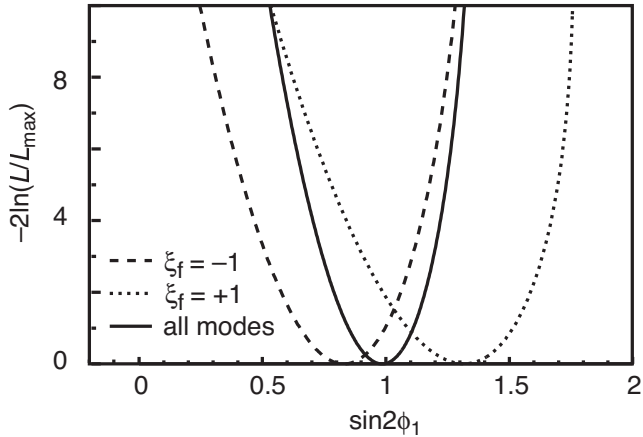}
  }
 \end{center}
 \vspace{1ex}
 \caption{
 Values of $-2\ln(L/L_{\rm max})$ 
 {\em vs.}~$\sin 2\phi_1$ for the $\xi_f=-1$(dashed) and  $+1$(dotted) modes separately and 
 for both modes combined(solid). 
 }
 \label{fig:loglikeli}
\end{figure}
\begin{figure}[htbp]
 \begin{center}
  \resizebox{0.48\textwidth}{!}{
  \includegraphics{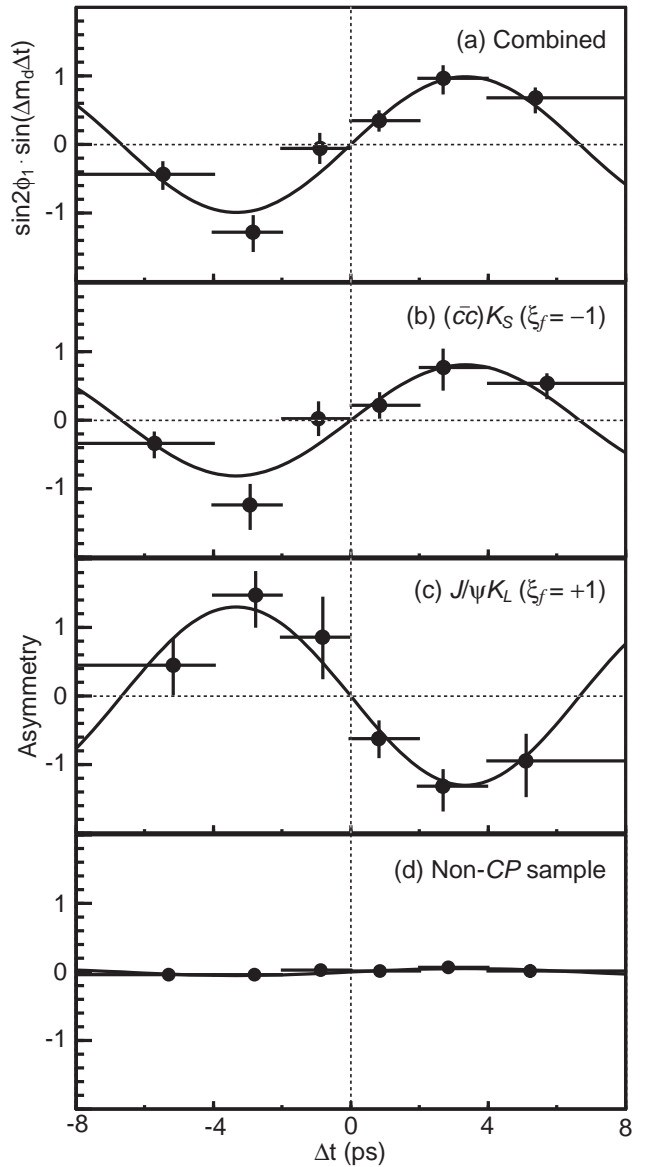}
  }
 \end{center}
 \vspace{1ex}
 \caption{
 (a) The asymmetry obtained
 from separate fits to each $\Delta t$ bin for 
 the full data sample; the curve is the result of 
 the global fit. The
 corresponding plots for the (b) $(c\overline{c})K_S^0$ ($\xi_f=-1$), (c) 
  $J/\psi K_L^0$ ($\xi_f = +1$) and (d) non-{\it CP} $B^0$ control samples
 are also shown.  The curves
 are the results of the combined fit applied separately to the
 individual data samples.
 }
 \label{fig:asym}
\end{figure}

We also checked the values of $\sin 2\phi_1$ in the different $r$ ranges of the
flavor tagging.  The results are listed in Table~\ref{cp_qdep}.
No systematic variation is seen.
Finally, we subdivided the $CP$ sample into three data taking 
periods: in 2000, from January to April 2001 and the rest.
The $\sin 2\phi_1$ values we obtain are
$0.84 \pm 0.29$(stat), $1.33 \pm 0.30$(stat) and $0.94 \pm 0.20$(stat), respectively.
Again the results are consistent within the statistical fluctuations.

\begin{table}[htb]
\caption{ The $r$ dependence of $\sin 2\phi_1$ fit result. 
          Only statistical errors are shown.}
\medskip
\label{cp_qdep} 
\begin{tabular}{lcccc}
$r$ region    & 0.0-0.5  & 0.5-0.75  &  0.75-0.875  & 0.875-1.0   \\ \hline
Events        &  613  & 239  &  119 &  166  \\
$\sin 2\phi_1 $ & $0.60^{~+0.65}_{~-0.67}$ &  $0.40^{~+0.31}_{~-0.32}$
                & $1.56^{~+0.25}_{~-0.29}$ &  $1.05^{~+0.15}_{~-0.18}$ \\
\end{tabular}
\end{table}

\subsection{Systematic Errors}

The sources of systematic error we consider are listed in
Table~\ref{tbl:syst-all}\cite{footnoteSyserr}. 
Adding all the systematic errors in quadrature, we obtain
\begin{eqnarray*}
 \sin 2\phi_1 = 0.99 \pm 0.14({\rm stat}) \pm 0.06({\rm syst}).
\end{eqnarray*}
\begin{table}[hbt]
 \begin{center}
 \caption{List of systematic errors on $\sin 2\phi_1$.}
 \label{tbl:syst-all}
  \begin{tabular}{lcc} 
   source  &  $+$error  & $-$error \\
   \hline
   vertex reconstruction & +0.040 & $-$0.040 \\
   resolution function   & +0.022 & $-$0.032 \\
   wrong tag fraction & +0.022 & $-$0.025 \\
   physics ($\tau_{B^0}$, $\Delta m_d$, $J/\psi K^{*0}$) & +0.007 & $-$0.004 \\
   background fraction (except for $J/\psi K_L^0$) & +0.003 & $-$0.004 \\
   background fraction ($J/\psi K_L^0$) &  +0.020 & $-$0.020 \\
   background shape  &  +0.001 & $-$0.001 \\
   \hline
   total & +0.06 & $-$0.06 \\
  \end{tabular}
 \end{center}
\end{table}
Below we explain each item in order.
%
\subsubsection{Vertex reconstruction}
The largest contribution comes from vertex reconstruction.
We searched for possible biases 
by using two
different vertexing algorithms and changing the track selection criteria for the
tag-side vertex.  
In the alternative vertexing algorithm, we first obtain a seed 
vertex using tracks of good quality: an impact parameter from IP in $r$-$\phi$ 
direction is smaller than 2.5 times the $r$-$\phi$ vertex error;
the vertex error in $z$ is less than 0.5 mm; and the cms momenta are
larger than 0.3 GeV/$c$.  We then repeat the vertex fit using tracks
within 3 (4) $\sigma$ in $z$ from the seed vertex for
the cms track momentum less (larger) than 1 GeV/$c$, where $\sigma$
is the error of 
the seed vertex in $z$.
We also estimated the effects of the vertex resolution tails
using samples with small $\Delta t$ ($|\Delta t|<6$~ps)
and tighter vertex quality cuts.
%
\subsubsection{Resolution function}
We estimate the contribution due to the uncertainty in the resolution 
function by varying its parameters (given in Section \ref{sec:resfunc})
by $\pm 1\sigma$.
%
\subsubsection{Wrong tag fraction}
Systematic errors due to uncertainties
in the wrong tag fractions
given in Section \ref{sec:fbtgw} 
(Table~\ref{tag_sum}) 
are studied by varying wrong tag fraction individually
for each $r$ region. We added the contributions from each variation in quadrature.
%
\subsubsection{Physics parameters}
The $B$ meson lifetime and mixing parameter
are fixed to the world average 
values\cite{PDG2000} in our fit; i.e.
$ \tau_B=(1.548\pm0.032)$ ps
and
$ \Delta m_d = (0.472\pm0.017)$ ps$^{-1}$.
We estimate the systematic error by repeating the fit
varying these parameters by their errors.
Another physics-related uncertainty is the
$CP$ eigenvalue of $J/\psi K^{*0}$ ($\xi_{J/\psi K^{*0}}$)
measured
from the angular distribution of the decay daughters\cite{Itoh}.
This systematic uncertainty is determined 
from the $\pm 1\sigma$ uncertainty in the measurement.
%
\subsubsection{Background fraction except for $J/\psi K_L^0$}
The background fraction in our pdf, $1-f_{\rm sig}$, is calculated from
the signal and background distribution functions of $\Delta E$ and $M_{\rm bc}$ 
as described in Section ~\ref{sec:signalprob}.
The distribution functions of $\Delta E$ and $M_{\rm bc}$ are
determined from data or the MC simulation depending on the decay mode.
To estimate the systematic errors associated with the choice of
parameterization, we varied the parameters obtained from the MC simulation by
$\pm 2\sigma$ and the parameters obtained from the data by
$\pm 1\sigma$. The likelihood fit was repeated.
A wider range of uncertainty was conservatively chosen for
parameters obtained from the MC simulation to take into account the possible
difference between the MC simulation and data.
We also estimated the systematic errors for the integrated background fractions,
listed in Table~\ref{bkg_frac}, by varying these parameters
by $\pm 1\sigma$.
We added the results of these calculations for each decay mode in
quadrature.
%
\subsubsection{Background fraction for $J/\psi K_L^0$}
As described in Section \ref{sec:psikl_rec},
the background fraction for the $J/\psi K_L^0$ sample
is obtained from a fit to $p^{\rm cms}_B$ distribution 
and is given in Table~\ref{fitresult}.
In this fit, the sum of components is automatically
constrained to the total number of events in the signal region.
Thus, the signal yield and the size of other backgrounds are
strongly anti-correlated.
To determine the systematic error on $\sin 2\phi_1$ that comes
from the uncertainty of the background, we need to take this anti-correlation
into account.
To this end, we repeat the fit to the $p^{\rm cms}_B$
distribution with the background fractions as free parameters but
with the signal yield fixed +1$\sigma$ or
$-$1$\sigma$ away from the central value obtained in the nominal fit.
The resultant background yields are used to repeat the procedure
to obtain $\sin 2\phi_1$. We regard the difference between
the thus obtained value and our nominal $\sin 2\phi_1$ value
as the systematic error.
We also check the systematic error due to the uncertainty in the \textit{CP} content of the background.
We repeated the fit varying parameters to determine the various background fractions.
Since these parameters are obtained from the MC simulation, we estimate the
systematic error by conservatively changing each parameter 
by $\pm 2\sigma$ and adding the resulting changes
in quadrature.
%
\subsubsection{Background shape}
The parameters that determine ${\cal P}_{\rm bkg}(\Delta t)$ and $R_{\rm bkg}$, 
given in Section \ref{sec:rbkg}, are varied within their errors and fits are
repeated.

\subsection{Cross checks }
We performed several cross checks:
The fitting procedure was examined using MC samples based on
our likelihood functions (toy MC samples).
We also measured the $B$ meson lifetime
using the same vertex reconstruction method.
In addition we tested non-$CP$ control samples. 
These cross checks are described below.

\subsubsection{Ensemble test}
A thousand toy MC
samples, each containing 1137 events,  are generated
based on our likelihood function to check the fitting procedure.
Figure~\ref{fig:toy_mc} shows the distribution of residuals
($\sin 2\phi_1$(fit) $- \sin 2\phi_1$(input)),
pulls (residuals divided by fit errors),
and the positive and negative errors on  
$\sin 2\phi_1$ returned from the fits.
All the toy MC samples have an input value of $\sin 2\phi_1=0.99$.
The center of the residual in Fig.~\ref{fig:toy_mc}~(a)
is consistent with zero, and the standard deviation of
the pull distribution in Fig.~\ref{fig:toy_mc}~(b) is consistent with unity.
Therefore the global fit returns the input $\sin 2\phi_1$ value and a reasonable error.
Figure~\ref{fig:toy_mc} (c) and (d) also show that
the positive and negative errors obtained from the fit are consistent with
expectations.

\begin{figure}[tbhp]
\begin{center}
\resizebox{0.48\textwidth}{!}{\includegraphics{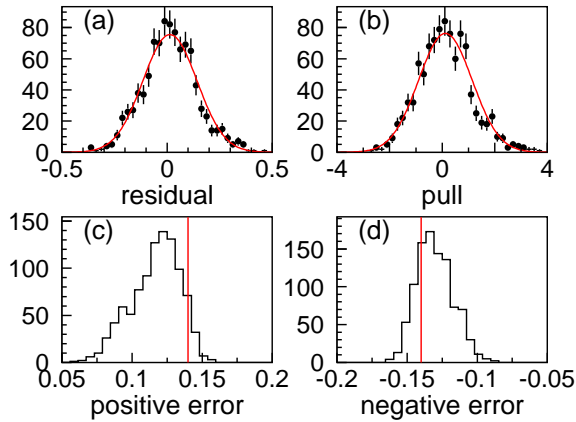}}
\vspace{1ex}
\caption{
The result of our toy Monte Carlo test of our fit: 
(a) the distribution of residuals
 (~$\sin 2\phi_1$(fit) $- \sin 2\phi_1$(input)~), 
(b) pull distribution (~(residual)/(error of fit)~),
 and (c) the positive and (d) negative errors of 
 $\sin 2\phi_1$ returned from the fits. The vertical lines in
figures (c) and (d) indicate the errors obtained from the fit to data.
}
\label{fig:toy_mc}
\end{center}
\end{figure}
%
We also generate toy MC samples for $J/\psi K^0_S$ 
and
$J/\psi K^0_L$. 
The average negative errors are 0.19 and 0.28 for $J/\psi K^0_S$ and
$J/\psi K^0_L$, respectively. Our measurement for $J/\psi K^0_S$ (0.20) is
in good agreement with the expectation, while the result for
for $J/\psi K^0_L$  (0.23) is smaller.
We obtain the probability of obtaining smaller errors than this
measurement to be 1.4\% for $J/\psi K^0_L$, which is within
a possible fluctuation.

\subsubsection{$B^0$ lifetime}
\label{subsubsec:b_lifetimes}
The $B^0$ lifetime
has been measured with the same data sample.  
We apply the same vertex reconstruction algorithm for fully reconstructed
$B^0$ decays as for the $CP$ decays and the tracks on the tag-side.
Unbinned maximum likelihood fits are made with an exponential pdf convolved
with the same $\Delta t$ resolution function and background pdf as in
the fit for $CP$ eigenstates.
For the combined $B^0 \to J/\psi K^{*0}(K^+\pi^-)$, $D^{(*)-}\pi^+$, $D^{*-}\rho^+$
and $D^{*-}\ell^+\nu$ decay modes, the $B^0$ lifetime is measured to 
be $\tau_{B^0} = 1.547 \pm 0.021({\rm stat.})$~ps.
The result is consistent with the world average value\cite{PDG2000}.

\subsubsection{Tests on control samples }  \label{sec:contrl}
We use control samples of non-$CP$ eigenstates,
$B^0\rightarrow J/\psi K^*(K^+\pi^-)$, $B^0 \rightarrow D^{(*)-}\pi^+$,
$B^0 \rightarrow D^{*-}\rho^+$ and
$B^0 \rightarrow D^{*-}\ell^+\nu$,
to check for biases in the analysis.
We perform the same fit to these control samples as for the $CP$-eigenstate modes.
The results, summarized in Table~\ref{cp_control},
show no systematic tendency.
A combined fit to all the modes yields $0.05 \pm 0.04$, consistent
with zero at the $1.2\sigma$ level, as shown in Fig.~\ref{fig:asym}~(d).

\begin{table}[htb]
\caption{$CP$ asymmetries for control samples.}
\medskip
\label{cp_control}
\begin{tabular}{lccc}
            & $J/\psi K^{*0}(K^+\pi^-)$ & 
              $D^{(*)-}\pi^+$ & 
              $D^{*-}\ell^+\nu$  \\
 & & and $D^{*-}\rho^+$ & \\
\hline
Events &   816  &  5560  &  10232  \\
$CP$ asymmetry  &  $0.01\pm 0.14 $ &
        $ 0.12\pm 0.06 $  &  $0.01 \pm 0.05$ \\
\end{tabular}
\end{table}

We check for a possible bias due to $CP$ asymmetry in the background.
We fit $J/\psi K_L^0$ candidates in the background region (1.0 $< p^{\rm cms}_B <$ 2.0 GeV/$c$)
treating all the events as $J/\psi K_L^0$ candidates.
Note that the fraction of events with definite $CP$ in this region of $p^{\rm cms}_B$ is expected to be negligible.
The result is $\sin2\phi_1 = 0.49~\pm0.35$, consistent with zero at the 1.4$\sigma$ level.

\subsection{Discussion}
\label{sec:discussions}
We have performed several statistical analyses of the results described in
the previous sections.
  Using a Gaussian likelihood function based on the statistical and 
  the systematic errors,
  we calculated the confidence intervals bounded
  by the physical region for $\sin 2 \phi_1$ using two methods:
  the Feldman-Cousins\cite{FELCOUS} frequentist approach and
  the Bayesian method with a flat prior pdf.
 We find a lower bound on $\sin 2 \phi_1$ of 0.70 at the 95\% C.L. in both cases.
  We also estimated the Bayesian lower limit using the exact 
  likelihood function, shown in Fig.~\ref{fig:loglikeli}, and obtained 0.69.
 We conclude that the likelihood function is Gaussian to
   a good approximation. 
Combinations of indirect measurements typically constrain
$0.50 < \sin 2\phi_1 < 0.86$
in the framework of the SM\cite{HOECKER2001}.
Although our measured value is large, it is consistent
with the higher range of the SM prediction. We are continuing
the measurement
with much higher statistics in order to test the KM ansatz more precisely.

Finally we comment on the possibility of direct $CP$ violation.
The signal pdf for a neutral $B$ meson decaying into a $CP$ eigenstate 
(Eqs.~\ref{pdf_sig_1} and \ref{pdf_sig_2})
can be expressed 
in a more general form as
\begin{eqnarray}
\label{eq:deltat_general}
&{\cal P}_{\rm sig}&(\Delta t,q,w_l,\xi_f) =
 \frac{ e^{-|\Delta t|/\tau_{B^0}} }{2\tau_{B^0}(1+|\lambda|^2)} \times
 \nonumber \\
 & & \left\{
 {1+|\lambda|^2 \over 2} - q(1-2w_l) [\xi_f {\cal A} |\lambda| \sin(\Delta m_d\Delta t) 
\right.
\nonumber \\
&& \left.
    - {|\lambda|^2 -1 \over 2} \cos(\Delta m_d\Delta t) ]
 \right\},
\end{eqnarray}
where $\lambda$ is a complex parameter 
that depends on both
$B^0 \bzb$ mixing and on the amplitudes for $B^0$ and $\bzb$ decay
to a $CP$ eigenstate\cite{BCP}.
The coefficient of $\sin(\Delta m_d\Delta t)$ is given by
${\cal A}|\lambda| = -\xi_f Im\lambda$
and is equal to $\sin 2 \phi_1$ in the SM. 
The presence of the cosine term
($|\lambda| \neq 1$) would indicate direct $CP$ violation.
Throughout our study we have assumed
$|\lambda| = 1$,  
as expected in the SM\cite{BCP}.
In order to test this assumption,
we also performed a fit using the above expression with 
${\cal A}$
and $|\lambda|$ as free parameters,
keeping everything else the same.
We obtain  
  $|\lambda| = 1.09 \pm 0.14$ and ${\cal A} = 0.80 \pm 0.19$
for the $J/\psi K_S^0(\pi^+\pi^-)$ sample only, and
  $|\lambda| = 1.03 \pm 0.09$ and ${\cal A} = 0.99 \pm 0.14$
for all $CP$ modes combined, where the errors are statistical 
only. This result confirms the assumption used in our analysis.

\section{Conclusion}
\label{sec:conclusion}

We have measured  the $CP$ violation parameter $\sin 2\phi_1$
at the KEKB asymmetric $e^+e^-$ collider
using a data sample of 29.1~fb$^{-1}$ 
recorded on the $\Upsilon(4S)$ resonance with the Belle detector.
To extract $\sin 2\phi_1$, we apply a maximum likelihood fit to
the 1137 candidate $B$ meson decays to $CP$ eigenstates.
We obtain
\[
 \sin 2\phi_1 = 0.99 \pm 0.14({\rm stat}) \pm 0.06({\rm syst}).
 \]
We conclude that there is large $CP$
violation in the neutral $B$ meson system.  
A zero value for
$\sin 2\phi_1$ is ruled out by more than six standard deviations.

We have calculated the confidence intervals bounded
  by the $\sin 2 \phi_1$ physical region for the Bayesian and
  Feldman-Cousins frequentist approaches. The lower bound is
  found to be 0.70 at the 95\% C.L. in both cases.
Our result is consistent
with the higher range of values allowed
from the constraints of the Standard Model
as well as with our previous measurements\cite{BELLE_cp1}.

\acknowledgements
We wish to thank the KEKB accelerator group for the excellent
operation of the KEKB accelerator.
We acknowledge support from the Ministry of Education,
Culture, Sports, Science, and Technology of Japan
and the Japan Society for the Promotion of Science;
the Australian Research Council
and the Australian Department of Industry, Science and Resources;
the National Science Foundation of China under contract No.~10175071;
the Department of Science and Technology of India;
the BK21 program of the Ministry of Education of Korea
and the CHEP SRC program of the Korea Science and Engineering Foundation;
the Polish State Committee for Scientific Research
under contract No.~2P03B 17017;
the Ministry of Science and Technology of the Russian Federation;
the Ministry of Education, Science and Sport of the Republic of Slovenia;
the National Science Council and the Ministry of Education of Taiwan;
and the U.S.\ Department of Energy.


\begin{thebibliography}{10}
\bibitem{CC}
Throughout this report, when a mode is quoted the inclusion of the charge
  conjugate mode is implied.
\bibitem{footnoteJPSIKSTAR}
For $J/\psi K^*$ candidates, we adopt slightly different windows. We require
  $2.95 \le M_{e^+e^-} \le 3.15 {\rm GeV}/c^2$ and $3.05 \le M_{\mu^+\mu^-} \le
  3.15 {\rm GeV}/c^2$.
\bibitem{footnoteKLPB3}
If there is more than one pion candidate, we select the one with the invariant
  mass of $K_L$ and the pion closest to the nominal $K^*$ mass.
\bibitem{footnoteMultiCand}
If two or more combinations of $J/\psi$ and $K_L$ clusters are found, the one
  that gives the best $p^{\rm cms}_B$ is selected.
\bibitem{footnoteDSTARLNU}
For a precise estimation of the uncorrelated background,
we require that a $B^0 \to D^{*-} \ell^+ \nu$ candidate
should not be in the signal region when
we artificially flip the
lepton momentum vector.
\bibitem{footnoteKstar}
The result for $J/\psi K^{*0}(K_S\pi^0)$ quoted in Ref.\cite{BELLE_cp2} was
  updated with an improved treatment of the background fraction.
\bibitem{footnoteSyserr}
The errors originating from the flavor tagging and from the resolution function
  differ slightly, after a further investigation, from the values quoted in
  Ref.~\cite{BELLE_cp2}. The total systematic error is unchanged.
\end{thebibliography}
\end{document}